\documentclass[aps,prb,twocolumn,amsmath,amssymb]{revtex4}

\usepackage{graphicx}
\usepackage{bm}
\usepackage{bbold}
\usepackage{hhline}
\usepackage{amsmath}
\usepackage{amssymb}

\def\bk{ \bm{k} }
\def\br{ \bm{r} }
\def\bq{ \bm{q} }
\def\bp{ \bm{p} }
\def\bmv{ \bm{v} }

\def\bmd{ \bm{d} }

\def\bmeta{ \bm{\eta} }
\def\hbms{ \hat{\bm{\sigma}} }
\def\calF{ {\cal F} }
\def\re{ \,\mathrm{Re}\, }
\def\im{ \,\mathrm{Im}\, }
\def\tr{ \mathrm{tr} }
\def\Tr{ \mathrm{Tr} }
\def\Eb{ {\cal E}_b } 
\def\sign{ \mathrm{sign} }

\begin{document}
\title{Ginzburg-Landau energy of multiband superconductors with interband pairing}

\author{K. V. Samokhin\footnote{E-mail: kirill.samokhin@brocku.ca}}
\affiliation{Department of Physics, Brock University, St. Catharines, Ontario L2S 3A1, Canada}


\begin{abstract}
We derive microscopically the Ginzburg-Landau free energy functional for a superconductor in which the Cooper pairs are formed not only by quasiparticles from the same band, but also by quasiparticles from different bands. In the simplest case of an $s$-wave or $d$-wave pairing in a two-band system, the order parameter has three components describing two intraband and one interband pair condensates. The interband pairing-specific terms in the free energy bring about some qualitative changes in the phase diagram, for example, time-reversal symmetry breaking superconducting states are stabilized at low temperatures.  
\end{abstract}

\maketitle

\section{Introduction}
\label{sec: Intro}

The properties of multiband, in particular two-band, superconductors (SCs) have been at the forefront of condensed matter physics research since the discovery of superconductivity in MgB$_2$
(Refs. \onlinecite{Naga01} and \onlinecite{BC15}). The list of materials in which multiband effects are thought to play an important role also includes Sr$_2$RuO$_4$ (Refs. \onlinecite{MM03} and \onlinecite{KB09}), 
NbSe$_2$ (Ref. \onlinecite{Boak03}), various heavy-fermion compounds,\cite{Bauer04,Tan05} iron-based SCs,\cite{Norm08,Hirsch11} doped topological insulators,\cite{Wray11,Fu10} superconducting oxide interfaces,\cite{Trevi18,Singh22} 
and others.

In the simplest theoretical approach,\cite{Suhl59,Mosk59} the Bardeen-Cooper-Schrieffer (BCS) model is extended to the two-band case by assuming that the pairing interaction shells near the 
Fermi surfaces in the two bands do not overlap, so that the Cooper pairs are formed only by the same-band quasiparticles. In this case, the order parameter in a one-dimensional (1D) pairing channel, 
e.g., $s$-wave or $d$-wave, has two components, $\eta_1$ and $\eta_2$, which describe the pair condensates in the two bands. The Cooper pairs can scatter from one band to the other producing a ``Josephson coupling'' 
between the bands, which depends on the relative phase of the two condensates. It is this coupling that gives rise to the most spectacular differences from the single-band case, such as the Leggett modes,\cite{Legg66,Shar02} phase solitons,\cite{Tanaka01} and fractional vortices,\cite{Baba02} see reviews in Refs. \onlinecite{Lin14} and \onlinecite{Tanaka15}.

The recent experimental developments have motivated a further extension of the theory of multiband superconductivity in which the pairing of quasiparticles from different bands is taken into account. 
The interband Cooper pairing can be incorporated into the BCS framework by assuming that the pairing interaction energy cutoff $\epsilon_c$ exceeds the band splitting. For realistic band structures that would likely require a non-phononic mechanism of pairing. 
In an alternative approach, one starts with a real-space pairing between different atomic orbitals in a crystal lattice and then transforms it into the band representation, which in general produces both intraband and interband pairing terms in the Hamiltonian.\cite{Moreo09,Fisch13,Ram16,Nomoto16,Nica17} 

Assuming that the interband pairs are created through one or another microscopic mechanism, one can use the group theory to classify the possible symmetries 
of the intraband and interband gap functions. 
Such phenomenological approach has proved to be very useful in the studies of fermionic superfuilds and superconductors,\cite{VG85,SU-review,TheBook} allowing one to determine the stable states and possible structures of the gap nodes even if the microscopic pairing mechanism is not known.  

In this paper, we derive the Ginzburg-Landau (GL) free energy functional for a multiband superconductor from a microscopic theory. We assume that there are two bands participating in superconductivity and 
take into account both intraband and interband pairing. Our calculations are based on an extended BCS model, in which the pairing shell in the momentum space contains both Fermi surfaces. Although we mostly focus on 1D pairing channels in a tetragonal SC, which correspond to 1D irreducible representations of the crystal point group $\mathbf{D}_{4h}$, 
our approach can be straightforwardly generalized to other crystal symmetries, higher-dimensional representations, and any number of bands. The condensate of the Cooper pairs formed by the quasiparticles from different bands is described by an additional order parameter component $\tilde\eta$. Therefore, the GL free energy depends on the three-component order parameter $\bmeta=(\eta_1,\eta_2,\tilde\eta)$. This leads to a more complicated structure of the free energy and a number of novel features, compared to the intraband-only GL theory.

The paper is organized as follows. In Sec. \ref{sec: Gap symmetry}, we review the symmetry-based classification of the intraband and interband gap functions and show, in particular, that the latter depend crucially on the symmetries of the two Bloch bands involved in the pairing. In Sec. \ref{sec: H-int-symmetry}, the symmetry analysis is applied to a generalized BCS Hamiltonian including all possible intraband and interband pairing
interactions. In Sec. \ref{sec: GL functional}, the GL functional is derived for the order parameter $\bmeta$, which has two intraband and one interband components. 
Some of the effects brought about by the interband pairing, namely, the emergence of stable time-reversal (TR) symmetry-breaking states in a two-band SC, are discussed in Sec. \ref{sec: stable states}. 
Throughout the paper we use the units in which $\hbar=k_B=1$, neglecting, in particular, the difference between the quasiparticle wave vector and momentum.

\section{Pairing symmetry: General analysis}
\label{sec: Gap symmetry}

We consider a centrosymmetric TR-invariant crystal described by the point group $\mathbb{G}$. External fields and impurities are neglected. The exact band states $|\bk,n,s\rangle$, which 
incorporate all effects of the periodic crystal lattice potential and the electron-lattice
spin-orbit (SO) coupling, are twofold degenerate at each wave vector $\bk$ due to the combined symmetry ${\cal C}=KI$, called conjugation,\cite{Kittel-book} where $K$ is the TR operation and $I$ is the spatial inversion.
We use the index $n$ to label the bands and also an additional index $s=1,2$ 
to distinguish two orthonormal Bloch states, $|\bk,n,1\rangle$ and $|\bk,n,2\rangle\equiv{\cal C}|\bk,n,1\rangle$, within the same band. In the presence of the SO coupling, the Bloch states have both spin-up 
and spin-down components, and $s$, called the Kramers index or the conjugation index, is not the same as the electron spin projection. 

The Bloch bands are classified according to the irreducible double-valued corepresentations (coreps) of the magnetic point group $\mathbb{G}+{\cal C}\mathbb{G}$ at the $\Gamma$ point, see Appendix \ref{app: Bloch-states}. 
In a given band, the electron creation operators in the Bloch states transform under the point-group operations and TR in the following way:\cite{Sam19-PRB}
\begin{equation}
\label{c-transform-g}
  gc^\dagger_{\bk,ns}g^{-1}=\sum_{s'}c^\dagger_{g\bk,ns'}{\cal D}_{n,s's}(g),\quad g\in\mathbb{G},
\end{equation}
and
\begin{equation}
\label{tilde c-c}
  \tilde c^\dagger_{\bk,ns}\equiv Kc^\dagger_{\bk,ns}K^{-1}=p_n\sum_{s'}c^\dagger_{-\bk,ns'}(-i\hat\sigma_y)_{s's}.
\end{equation}
Here $\hat{\cal D}_n(g)$ is the $\Gamma$-point corep matrix in the $n$th band and $p_n=\pm 1$ is the band parity.
We use the notation $\hat\sigma_0$ and $\hat{\bm\sigma}=(\hat\sigma_x,\hat\sigma_y,\hat\sigma_z)$ respectively for the identity matrix and the Pauli matrices in the Kramers space. 

In this paper, we consider only the point group $\mathbb{G}=\mathbf{D}_{4h}$, which describes the symmetry of numerous important superconductors,
from the high-$T_c$ cuprates and iron pnictides to Sr$_2$RuO$_4$. Due to the presence of the inversion symmetry,
the $\Gamma$-point coreps are either inversion-even ($\Gamma^+$) or inversion-odd ($\Gamma^-$). 
The magnetic group $\mathbf{D}_{4h}+{\cal C}\mathbf{D}_{4h}$ has four double-valued coreps, $\Gamma_6^\pm$ and $\Gamma_7^\pm$, only $\Gamma_6^+$ being equivalent 
to the spin-$1/2$ corep.\cite{Lax-book,BC-book} Therefore, $\Gamma_6^+$ bands are pseudospin bands, while $\Gamma_6^-$ and $\Gamma_7^\pm$ bands are non-pseudospin bands. 

Suppose there are two bands crossing the chemical potential and participating in superconductivity. The bands can have the same or different symmetries, i.e., correspond to the same or different $\Gamma$-point coreps. In the $\mathbf{D}_{4h}$ case, there are ten possible two-band combinations: $(\Gamma_6^+,\Gamma_6^+)$, $(\Gamma_6^+,\Gamma_6^-)$, $(\Gamma_6^+,\Gamma_7^+)$, etc.
At the mean-field level, the Hamiltonian in a uniform superconducting state has the form $\hat H_{MF}=\hat H_0+\hat H_{sc}$, where
\begin{equation}
\label{H0-gen}
  \hat H_0=\sum_{n=1,2}\sum_{\bk, s}\xi_n(\bk)c^\dagger_{\bk,ns}c_{\bk,ns}
\end{equation}
describes non-interacting quasiparticles in two twofold degenerate Bloch bands. The band dispersions $\xi_n(\bk)=\xi_n(-\bk)$ are counted from the chemical potential, which is set equal to the Fermi energy $\epsilon_F$. 
Without loss of generality, we assume that $\xi_1(\bk)<\xi_2(\bk)$ at all $\bk$. The pairing Hamiltonian is given by
\begin{eqnarray}
\label{H-mean-field-general}
  \hat H_{sc} = \frac{1}{2}\sum_{nn'}\sum_{\bk,ss'}\Delta_{nn',ss'}(\bk)c^\dagger_{\bk,ns}\tilde c^\dagger_{\bk, n's'}+\mathrm{H.c.},
\end{eqnarray}
where the operators $\tilde c^\dagger_{\bk,ns}$ create electrons in TR-transformed states, see Eq. (\ref{tilde c-c}). The intraband pairing in the $n$th band is described by the gap functions $\hat\Delta_{nn}$, whereas $\hat\Delta_{12}$ 
and $\hat\Delta_{21}$ describe the pairing of quasiparticles from different bands (the interband pairing). The latter can be included in a general mean-field model on the same footing as the
intraband gap functions. Microscopically, interband pairs appear in a BCS-like model if the pairing interaction shells near the Fermi surfaces, which are defined by $|\xi_1|,|\xi_2|\leq\epsilon_c$, overlap, 
i.e., if the pairing interaction energy cutoff $\epsilon_c$ exceeds the interband splitting, see Sec. \ref{sec: H-int-symmetry}.

For each pair of bands, the gap function is a $2\times 2$ matrix in the Kramers space, which can be represented as follows: 
\begin{equation}
\label{Delta-s-t-expansion-general}
  \hat\Delta_{nn'}(\bk)=\psi_{nn'}(\bk)\hat\sigma_0+\bmd_{nn'}(\bk)\hat{\bm\sigma}.
\end{equation}
By analogy with the standard (single-band) theory of superconductivity, see, for instance, Refs. \onlinecite{SU-review} and \onlinecite{TheBook}, 
one can call $\psi_{nn'}$ and $\bmd_{nn'}$ the singlet and triplet components, respectively. Note that the factors $i\hat\sigma_y$ are absent from the expression (\ref{Delta-s-t-expansion-general}), 
because the gap functions are defined in Eq. (\ref{H-mean-field-general}) 
as the measures of the pairing between the quasiparticles in the states $|\bk,n,s\rangle$ and $K|\bk,n',s'\rangle$, not in $|\bk,n,s\rangle$ and $|-\bk,n',s'\rangle$. 
This ensures\cite{Blount85} that the Bogoliubov-de Gennes Hamiltonian is a proper first-quantization Hamiltonian and that the gap functions have natural transformation properties under the symmetry operations. 
The anticommutation of the fermionic operators imposes the following constraint:
\begin{equation}
\label{Delta-constraint-anticommutation}
  \hat\Delta_{nn'}(\bk)=p_np_{n'}\hat\sigma_y\hat\Delta^\top_{n'n}(-\bk)\hat\sigma_y,
\end{equation} 
therefore, $\psi_{nn'}(\bk)=p_np_{n'}\psi_{n'n}(-\bk)$ and $\bmd_{nn'}(\bk)=-p_np_{n'}\bmd_{n'n}(-\bk)$. We see that, while the intraband singlet (triplet) gap functions are necessarily even (odd) in $\bk$, the parity of the interband pairing is not determined by the anticommutation requirement alone.

Applying Eq. (\ref{c-transform-g}) to the pairing Hamiltonian (\ref{H-mean-field-general}), we find that the symmetry operations from the point group induce the following transformation of the gap functions:
\begin{equation}
\label{Delta-transform-g}
  g:\ \hat\Delta_{nn'}(\bk)\to\hat{\cal D}_n(g)\hat\Delta_{nn'}(g^{-1}\bk)\hat{\cal D}^\dagger_{n'}(g).
\end{equation}
Thus, the gap transformation properties are nonuniversal, in the sense that they depend on the symmetries of the bands involved in the pairing. The singlet components $\psi_{nn'}$ do not necessarily  
tranform as scalar functions of $\bk$, while the triplet components $\bmd_{nn'}$ are not always pseudovectors. Even the intraband pairing may be affected: it was shown in Ref. \onlinecite{Sam19-PRB} that in certain bands in trigonal and hexagonal superconductors the standard classification of triplet pairing states breaks down, with profound consequences for the gap nodal structure. 
Regarding the response of the gap functions to TR, it follows from Eq. (\ref{tilde c-c}) that
\begin{equation}
\label{Delta-transform-K}
  K:\ \hat\Delta_{nn'}(\bk)\to\hat\Delta^\dagger_{n'n}(\bk).
\end{equation}
To obtain this, we used the antilinearity of the TR operator and the fact that $K\tilde c^\dagger_{\bk,ns}K^{-1}=-c^\dagger_{\bk,ns}$.

\subsection{Order parameter components}
\label{sec: gap expansion}

According to the Landau theory of phase transitions, the gap functions, both intraband and interband, must correspond to the same single-valued irreducible representation (irrep) $\gamma$ of the point group $\mathbb{G}$, which is called the pairing channel. For $\mathbb{G}=\mathbf{D}_{4h}$, there are ten single-valued irreps of either parity, eight 1D and two two-dimensional (2D), 
see Ref. \onlinecite{Lax-book}. In particular, the 1D irrep $A_{1g}$ describes the ``$s$-wave'' pairing, whereas the 2D irrep $E_u$ describes the ``$p$-wave'' pairing.
Note that we use the ``chemical'' notation for the single-valued irreps corresponding to the pairing channels, reserving the $\Gamma$ notation for the double-valued coreps describing the symmetry of the Bloch bands.

For each pair of bands, the gap function can be represented as a linear combination of the matrix basis functions of the $d$-dimensional irrep $\gamma$ as follows:
\begin{equation}
\label{Delta-expansion-general}
  \hat\Delta_{nn'}(\bk)=\sum_{a=1}^d\eta^a_{nn'}\hat\phi_{nn'}^{a}(\bk).
\end{equation}
The expansion coefficients $\eta^a_{nn'}$ here play the role of the order parameter components and are found by minimizing the free energy of the superconductor. Transformation of the $2\times 2$ matrix basis functions $\hat\phi_{nn'}^{a}(\bk)$ under the point group operations follows immediately from Eq. (\ref{Delta-transform-g}): 
\begin{eqnarray}
\label{phi-transform-g}
  g:\hat\phi_{nn'}^{a}(\bk) &\to & \hat{\cal D}_n(g)\hat\phi_{nn'}^{a}(g^{-1}\bk)\hat{\cal D}^\dagger_{n'}(g) \nonumber\\
  &=& \sum_{b=1}^d\hat\phi_{nn'}^{b}(\bk)D_{\gamma,ba}(g),
\end{eqnarray}
where $\hat D_\gamma(g)$ is the $d\times d$ representation matrix. In particular, the basis functions in a 1D pairing channel satisfy the following equation:
\begin{equation}
\label{phi-equations-g-1D}
  \hat{\cal D}_n(g)\hat\phi_{nn'}(g^{-1}\bk)\hat{\cal D}^\dagger_{n'}(g)=\chi_\gamma(g)\hat\phi_{nn'}(\bk), 
\end{equation}
where $\chi_\gamma(g)$ is the character of $g$ in the irrep $\gamma$. Similarly to Eq. (\ref{Delta-s-t-expansion-general}), the basis functions can be represented as sums of the ``singlet'' and ``triplet'' components, the former 
containing the identity matrix $\hat\sigma_0$ and the latter -- the Pauli matrices $\hbms$.

Explicit expressions for the basis functions can be found by solving Eq. (\ref{phi-transform-g}), subject to several additional constraints. First, it follows from the anticommutation condition 
(\ref{Delta-constraint-anticommutation}) that
\begin{equation}
\label{phi-anticommutation-constraint}
  \hat\phi_{nn'}^a(\bk)=p_np_{n'}\hat\sigma_y\hat\phi_{n'n}^{a,\top}(-\bk)\hat\sigma_y.
\end{equation}
Second, our crystal has an inversion center, so we can put $g=I$ in Eq. (\ref{phi-transform-g}) and obtain:
\begin{equation}
\label{phi-transform-I}
  p_np_{n'}\hat\phi_{nn'}^{a}(-\bk)=P_\gamma\hat\phi_{nn'}^{a}(\bk),
\end{equation}
where $P_\gamma\equiv\chi_\gamma(I)=\pm 1$ is the parity of the pairing channel $\gamma$ (not to be confused with the band parities $p_1$ and $p_2$). 
Combining Eqs. (\ref{phi-anticommutation-constraint}) and (\ref{phi-transform-I}), we see that $\hat\phi_{nn'}^a(\bk)=P_\gamma\hat\sigma_y\hat\phi_{n'n}^{a,\top}(\bk)\hat\sigma_y$.
Therefore, the statement that an even pairing ($P_\gamma=+1$) is purely singlet, i.e., the basis functions contain only $\hat\sigma_0$, whereas an odd pairing ($P_\gamma=-1$) is purely triplet, i.e., 
the basis functions contain only $\hbms$, 
is true only for the intraband functions $\hat\phi_{nn}^a(\bk)$. For the interband gap functions, both the singlet and triplet components can be present simultaneously without violating the Pauli principle,
with the parity of $\hat\phi_{nn'}^{a}(\bk)$ determined by the relative parity of the bands, see the examples in Secs. \ref{sec: s-wave-D-4h} and \ref{sec: p-wave-D-4h} below. 

The final constraint on the basis functions is obtained using the response to TR. According to Eq. (\ref{Delta-transform-K}),
\begin{equation}
\label{phi-transform-K}
  K:\hat\phi_{nn'}^{a}(\bk)\to\hat\phi_{n'n}^{a,\dagger}(\bk).
\end{equation}
Note that $K^2=1$ when acting on the gap functions and the basis functions.
It follows from Eqs. (\ref{phi-transform-g}) and (\ref{phi-transform-K}) that, for a given pair of bands, the set $\{\hat\phi_{nn'}^{a}(\bk),\hat\phi_{n'n}^{a,\dagger}(\bk)\}$ with $a=1,...,d$ forms the basis of a $2d$-dimensional 
single-valued corep of the magnetic point group $\mathbb{G}+K\mathbb{G}$, which is derived from the irrep $\gamma$. The corep matrices are given by
$$
  \hat{\mathbb{D}}(g)=\left(\begin{array}{cc}
                    \hat D_\gamma(g) & 0 \\
                    0 & \hat D_\gamma^*(g)
                    \end{array}\right),\ 
  \hat{\mathbb{D}}(K)=\left(\begin{array}{cc}
                    0 & \hat{\mathbb{1}}_d \\
                    \hat{\mathbb{1}}_d & 0
                    \end{array}\right),
$$
where $\hat{\mathbb{1}}_d$ is the $d\times d$ unit matrix. According to Refs. \onlinecite{Lax-book} and \onlinecite{BC-book}, coreps or magnetic groups are classified into three cases, A, B, or C, which determine whether or not the TR symmetry leads to an additional degeneracy and also the type of this degeneracy.
In Case A, there is no additional degeneracy, i.e., the corep is reducible, whereas the TR symmetry brings about additional degeneracy of the ``doubling'' type in Case B and of the ``pairing'' type in Case C. 
One can show that all coreps for the point group $\mathbf{D}_{4h}$ are Case A. Therefore, the set of the TR-transformed basis functions 
$\hat\phi_{n'n}^{a,\dagger}(\bk)$ is the same as as the set of $\hat\phi^{a}_{nn'}(\bk)$, and one can put 
\begin{equation}
\label{K-constraint-basis-functions}
  \hat\phi^a_{nn'}(\bk)=\hat\phi_{n'n}^{a,\dagger}(\bk),
\end{equation}
for all pairs of bands. 

Returning to the order parameter components, it follows from Eqs. (\ref{Delta-constraint-anticommutation}) and (\ref{phi-anticommutation-constraint}) that
\begin{equation}
\label{eta-symmetric}
  \eta^a_{nn'}=\eta^a_{n'n}.
\end{equation}
Therefore, the superconducting state corresponding to a $d$-dimensional pairing channel in an $N$-band superconductor is described by $N(N+1)d/2$ independent order parameter components, of which $Nd$ characterize the intraband pair condensates and $N(N-1)d/2$ -- the interband ones. In 
the two-band case, the order parameter has $3d$ components: $\eta_{11}^a$, $\eta_{22}^a$, and $\eta_{12}^a$. 

Using Eqs. (\ref{Delta-transform-g}) and (\ref{phi-transform-g}), we see that under the point group operations 
the order parameter transforms as follows:
\begin{equation}
\label{eta-transform-g}
  g:\eta^a_{nn'}\to\sum_{b=1}^dD_{\gamma,ba}(g)\eta^b_{nn'}.
\end{equation}
This means that the structure of the GL free energy depends only on the pairing channel $\gamma$, but not on the symmetry of the electron bands participating in the pairing. The latter affects only the matrix structure and the momentum dependence 
of the basis functions. Finally, it follows from Eqs. (\ref{Delta-transform-K}) and (\ref{K-constraint-basis-functions}) that
\begin{equation}
\label{eta-transform-K}
  K:\eta^a_{nn'}\to\eta^{a,*}_{nn'},
\end{equation}
i.e., the action of TR on the order parameter is equivalent to complex conjugation. In a TR-invariant superconducting state, all components of the order parameter are real.

\subsection{Example: $s$-wave pairing}
\label{sec: s-wave-D-4h}

The $s$-wave pairing channel corresponds to the identity irrep $A_{1g}$. Here and below we assume a quasi-2D band structure, i.e., set $\bk=(k_x,k_y)$. 
The gap functions (\ref{Delta-expansion-general}) take the form $\hat\Delta_{nn'}(\bk)=\eta_{nn'}\hat\phi_{nn'}(\bk)$, where $n,n'=1,2$. 
The pairing channel is even ($P_\gamma=1$), therefore, according to Eqs. (\ref{phi-anticommutation-constraint}), 
(\ref{phi-transform-I}), and (\ref{K-constraint-basis-functions}), the intraband basis functions are given by $\hat\phi_{nn}(\bk)=\alpha_n(\bk)\hat\sigma_0$, where $\alpha_n$ are real and even in $\bk$. 
The interband basis functions can be sought in the form
\begin{equation}
\label{interband-phi-alpha-beta}
  \left.\begin{array}{c}
        \hat\phi_{12}(\bk)=\hat{\tilde\phi}(\bk)=\tilde\alpha(\bk)\hat\sigma_0+i\tilde{\bm{\beta}}(\bk)\hbms,\smallskip \\
	\hat\phi_{21}(\bk)=\hat{\tilde\phi}^\dagger(\bk)=\tilde\alpha(\bk)\hat\sigma_0-i\tilde{\bm{\beta}}(\bk)\hbms,
        \end{array}\right.
\end{equation}
where the real functions $\tilde\alpha$ and $\tilde{\bm{\beta}}$ are even (odd) in $\bk$ for the bands of the same (opposite) parity.

The momentum dependence of $\alpha_{1,2}$, $\tilde\alpha$, and $\tilde{\bm{\beta}}$ is found from the point-group constraint (\ref{phi-equations-g-1D}) with $\chi_\gamma(g)=1$ for all $g$, which should be solved for each pair of bands. 
The intraband basis functions are just real invariant scalars, satisfying $\alpha_n(g^{-1}\bk)=\alpha_n(\bk)$, so one can put $\alpha_{1,2}(\bk)=1$ without loss of generality.  
In contrast, the interband basis functions depend on the symmetries of the bands involved in the pairing and are listed
in Table \ref{table: phis-s-wave}, see Appendix \ref{app: s-wave-interband} for the details of the calculation. 

Introducing the shorthand notation
\begin{equation}
\label{s-wave-OP-psis}
  \eta_1\equiv\eta_{11},\quad \eta_2\equiv\eta_{22},\quad \tilde\eta\equiv\eta_{12}=\eta_{21},
\end{equation}
the intraband and interband gap functions take the following form:
\begin{equation}
\label{two-band-gaps-s-wave}
	\begin{array}{c}
	  \hat\Delta_{11}(\bk)=\eta_1\alpha_1(\bk)\hat\sigma_0,\quad \hat\Delta_{22}(\bk)=\eta_2\alpha_2(\bk)\hat\sigma_0,\smallskip \\
	  \hat\Delta_{12}(\bk)=\tilde\eta[\tilde\alpha(\bk)\hat\sigma_0+i\tilde{\bm{\beta}}(\bk)\hbms],\smallskip \\
	  \hat\Delta_{21}(\bk)=\tilde\eta[\tilde\alpha(\bk)\hat\sigma_0-i\tilde{\bm{\beta}}(\bk)\hbms].
	\end{array}
\end{equation}
The order parameter components $\eta_1$, $\eta_2$, and $\tilde\eta$ are found by minimizing the free energy of the superconductor, see Sec. \ref{sec: GL functional}.

We see from Eq. (\ref{two-band-gaps-s-wave}) that, while the structure of the intraband gap functions is standard for the singlet isotropic pairing, the interband gap functions exhibit unconventional features 
such as a nonzero triplet component and the parity which depends on the relative parity of the bands. For example, for the opposite-parity bands $(\Gamma_6^\pm,\Gamma_6^\mp)$ or $(\Gamma_7^\pm,\Gamma_7^\mp)$, 
we have
\begin{equation}
\label{Delta12-s-wave-G6G6}
  \hat\Delta_{12}(\bk)=i\tilde\eta(k_x\hat\sigma_x+k_y\hat\sigma_y),
\end{equation}
which looks like a $p$-wave gap function, but in fact remains invariant under all elements of the point group, i.e., correspons to the identity irrep $A_{1g}$. 
In particular, we have $I: \hat\Delta_{12}(\bk)\to p_1p_2\hat\Delta_{12}(-\bk)=\hat\Delta_{12}(\bk)$, according to Eq. (\ref{Delta-transform-g}).
The imaginary factor in $\hat\Delta_{12}$ ensures that $\tilde\eta\to\tilde\eta^*$ under the TR operation.

\begin{table}
\caption{Momentum dependence of the $s$-wave interband pairing in a quasi-2D crystal with $\mathbb{G}=\mathbf{D}_{4h}$ ($a$ is a real constant). First column: the $\Gamma$-point coreps of the bands 
participating in the pairing.}
\begin{tabular}{|ll|}
    \hline
      & $\hat{\tilde\phi}(\bk)$ \\ \hline
    $(\Gamma_6^\pm,\Gamma_6^\pm)$, $(\Gamma_7^\pm,\Gamma_7^\pm)$\qquad\qquad & $\hat\sigma_0+ak_xk_y(k_x^2-k_y^2)\hat\sigma_z$ \\ \hline
    $(\Gamma_6^\pm,\Gamma_6^\mp)$, $(\Gamma_7^\pm,\Gamma_7^\mp)$  &  $k_x\hat\sigma_x+k_y\hat\sigma_y$ \\ \hline
    $(\Gamma_6^\pm,\Gamma_7^\pm)$  &  $(k_x^2-k_y^2)\hat\sigma_0+ak_xk_y\hat\sigma_z$ \\ \hline
    $(\Gamma_6^\pm,\Gamma_7^\mp)$  &  $k_x\hat\sigma_x-k_y\hat\sigma_y$ \\ \hline
\end{tabular}
\label{table: phis-s-wave}
\end{table}

\subsection{Example: $p$-wave pairing}
\label{sec: p-wave-D-4h}

For the $p$-wave pairing channel, which corresponds to the 2D irrep $E_u$ of $\mathbf{D}_{4h}$, the gap functions (\ref{Delta-expansion-general}) 
take the form $\hat\Delta_{nn'}(\bk)=\sum_{a=1,2}\eta^a_{nn'}\hat\phi_{nn'}^{a}(\bk)$. 
The pairing channel is odd ($P_\gamma=-1$) and, according to Eqs. (\ref{phi-anticommutation-constraint}), 
(\ref{phi-transform-I}), and (\ref{K-constraint-basis-functions}), the intraband basis functions are given by $\hat\phi^a_{nn}(\bk)=\bm{\beta}^a_n(\bk)\hbms$, where $\bm{\beta}^a_n$ are real and odd in $\bk$. 
The interband basis functions can be sought in the form
$$
  \left.\begin{array}{c}
  \hat\phi_{12}^a(\bk)=\hat{\tilde\phi}_a(\bk)=i\tilde\alpha_a(\bk)\hat\sigma_0+\tilde{\bm{\beta}}_a(\bk)\hbms,\smallskip \\
  \hat\phi_{21}^a(\bk)=\hat{\tilde\phi}_a^{\dagger}(\bk)=-i\tilde\alpha_a(\bk)\hat\sigma_0+\tilde{\bm{\beta}}_a(\bk)\hbms,
  \end{array}\right. 
$$
where the real functions $\tilde\alpha_a$ and $\tilde{\bm{\beta}}_a$ are odd (even) in $\bk$ for the bands of the same (opposite) parity.

The momentum dependence of $\bm{\beta}^a_{1,2}$, $\tilde\alpha_a$, and $\tilde{\bm{\beta}}_a$ is found from Eq. (\ref{phi-transform-g}), which should be solved for each pair of bands. 
For the intraband basis functions, one can put $\bm{\beta}^1_n(\bk)=(0,0,k_x)$ and $\bm{\beta}^2_n(\bk)=(0,0,k_y)$.  
The interband basis functions depend on the symmetries of the bands involved in the pairing and are listed in Table \ref{table: phis-p-wave}. 

Introducing the shorthand notation
$$
  \eta_{1,a}\equiv\eta_{11}^a,\quad \eta_{2,a}\equiv\eta_{22}^a,\quad \tilde\eta_a\equiv\eta_{12}^a=\eta_{21}^a,
$$
where $a=1,2$, the intraband and interband gap functions corresponding to the $p$-wave pairing take the following form:
\begin{equation}
\label{two-band-gaps-p-wave}
	\begin{array}{c}
	  \hat\Delta_{11}(\bk)=(\eta_{1,1}k_x+\eta_{1,2}k_y)\hat\sigma_z,\smallskip \\ 
	  \hat\Delta_{22}(\bk)=(\eta_{2,1}k_x+\eta_{2,2}k_y)\hat\sigma_z,\smallskip \\
	  \hat\Delta_{12}(\bk)=\tilde\eta_1\hat{\tilde\phi}_1(\bk)+\tilde\eta_2\hat{\tilde\phi}_2(\bk),\smallskip \\ 
	  \hat\Delta_{21}(\bk)=\tilde\eta_1\hat{\tilde\phi}_1^\dagger(\bk)+\tilde\eta_2\hat{\tilde\phi}_2^\dagger(\bk).
	\end{array}
\end{equation}
The order parameter has six components: the intraband ones $\bm{\eta}_1=(\eta_{1,1},\eta_{1,2})$ and $\bm{\eta}_2=(\eta_{2,1},\eta_{2,2})$, and the interband ones $\bm{\tilde{\eta}}=(\tilde\eta_1,\tilde\eta_2)$, which can be found by minimizing the GL free energy of the superconductor. 

The structure of the intraband gap functions in Eq. (\ref{two-band-gaps-p-wave}) is standard for a quasi-2D $p$-wave pairing. In contrast, the interband gap functions look unusual, because they essentially depend 
on the symmetries of the bands and either contain a nonzero singlet component or are even in $\bk$. The latter possibility, namely,
$\hat\Delta_{12}(\bk)=\tilde\eta_1\hat\sigma_y-\tilde\eta_2\hat\sigma_x=\hat\Delta_{21}(\bk)$, is realized for any combination of the opposite-parity bands.

\begin{table}
\caption{Momentum dependence of the $p$-wave interband pairing in a quasi-2D crystal with $\mathbb{G}=\mathbf{D}_{4h}$ ($a$ is a real constant). First column: the $\Gamma$-point coreps of the bands 
participating in the pairing.}
\begin{tabular}{|ll|}
    \hline
      & $(\hat{\tilde\phi}_1(\bk),\;\hat{\tilde\phi}_2(\bk))$ \\ \hline
    $(\Gamma_6^\pm,\Gamma_6^\pm)$, $(\Gamma_7^\pm,\Gamma_7^\pm)$\quad\qquad & $(iak_y\hat\sigma_0+k_x\hat\sigma_z,\;-iak_x\hat\sigma_0+k_y\hat\sigma_z)$ \\ \hline
    $(\Gamma_6^\pm,\Gamma_6^\mp)$, $(\Gamma_7^\pm,\Gamma_7^\mp)$  &  $(\hat\sigma_y,-\hat\sigma_x)$ \\ \hline
    $(\Gamma_6^\pm,\Gamma_7^\pm)$  &  $(iak_y\hat\sigma_0+k_x\hat\sigma_z,\;-iak_x\hat\sigma_0+k_y\hat\sigma_z)$ \\ \hline
    $(\Gamma_6^\pm,\Gamma_7^\mp)$  &  $(\hat\sigma_y,-\hat\sigma_x)$ \\ \hline
\end{tabular}
\label{table: phis-p-wave}
\end{table}

\section{Full pairing Hamiltonian}
\label{sec: H-int-symmetry}

The symmetry analysis of the mean-field gap functions can be straightforwardly extended to the full Hamiltonian describing the pairing interaction in the basis of the exact band states. 
We have $\hat H=\hat H_0+\hat H_{int}$, where $\hat H_0$ is given by Eq. (\ref{H0-gen}) and
\begin{eqnarray}
\label{Hint-gen}
  && \hat H_{int} = \frac{1}{2{\cal V}}\sum\limits_{\bk\bk'\bq}\sum_{n_is_i}V_{s_1s_2s_3s_4}^{n_1n_2n_3n_4}(\bk,\bk';\bq) \nonumber\\
  && \qquad\times c^\dagger_{\bk+\frac{\bq}{2},n_1s_1}\tilde c^\dagger_{\bk-\frac{\bq}{2},n_2s_2}\tilde c_{\bk'-\frac{\bq}{2},n_3s_3}c_{\bk'+\frac{\bq}{2},n_4s_4}\qquad
\end{eqnarray}
is the pairing Hamiltonian, $n_i=1,2$ is the band index, and $s_i=1,2$ is the Kramers index. The Cooper pairing takes place between the quasiparticles in the states $|\bk+\bq/2,n_1,s_1\rangle$ and $K|\bk-\bq/2,n_2,s_2\rangle$, see Eq. (\ref{H-mean-field-general}). The center-of-mass momentum of the pairs is equal to $\bq$. Quasiparticles from different bands can form a pair with $\bq=\bm{0}$ if they have mismatched energies within the interaction energy shell, see below.

\begin{figure}
\includegraphics[width=5cm]{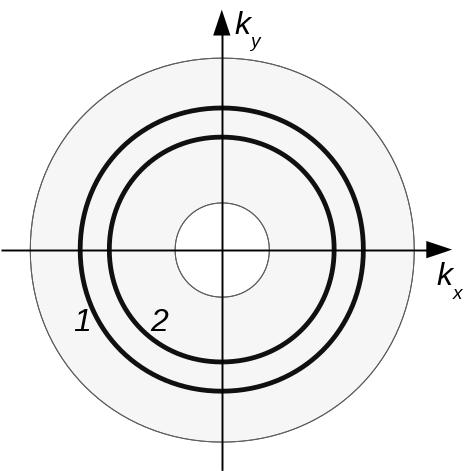}
\caption{The Fermi surfaces in the bands $1$ and $2$ within the BCS pairing shell (the shaded annulus).}
\label{fig: Fermi-surfaces}
\end{figure}

Guided by a considerable precedent in the theory of unconventional superconductivity,\cite{SU-review,TheBook} we make the following assumptions about the interaction function $V$. 
First, we neglect its dependence on the pair center-of-mass momentum $\bq$ (taking this dependence into account can lead to some interesting consequences, see Ref. \onlinecite{Sam13-LIs}, which are not considered here). 
Second, we assume, in the spirit of the BCS theory, that only the quasiparticles inside a pairing shell near the Fermi surface participate in the pairing. 
In the two-band case, the Fermi surface consists of two or more sheets corresponding to the solutions of the equations $\xi_1(\bk)=0$ and $\xi_2(\bk)=0$, and
\begin{eqnarray}
\label{V-BCS-cutoffs}
  \hat V^{n_1n_2n_3n_4}(\bk,\bk')\propto\theta(\epsilon_c-|\xi_{n_i}(\bk)|)\theta(\epsilon_c-|\xi_{n_2}(\bk)|)\nonumber\\
  \times \theta(\epsilon_c-|\xi_{n_3}(\bk')|)\theta(\epsilon_c-|\xi_{n_4}(\bk')|),
\end{eqnarray}
where $\theta(x)$ is the Heaviside step function and $\epsilon_c$ is the energy cutoff. Therefore, the interband pairing is present only if the BCS shells in the two bands overlap, i.e., if
$$
    \epsilon_c>\frac{\Eb}{2},\quad \Eb=\max_{\bk}|\xi_2(\bk)-\xi_1(\bk)|.
$$
The third assumption is that the momentum dependence of the pairing interaction inside the BCS shell can be represented in a factorized form:
\begin{eqnarray}
\label{V-factorized}
    && V_{s_1s_2s_3s_4}^{n_1n_2n_3n_4}(\bk,\bk') = -\frac{1}{2}V_{n_1n_2,n_3n_4}\nonumber\\
    && \qquad \times \sum_{a=1}^d \phi_{n_1n_2,s_1s_2}^{a}(\bk)\phi_{n_4n_3,s_4s_3}^{a,*}(\bk'),\qquad
\end{eqnarray}
where $V_{n_1n_2,n_3n_4}$ are the dimensional coupling constants and $\hat\phi^a_{nn'}(\bk)$ are the $2\times 2$ matrix basis functions of the $d$-dimensional pairing channel $\gamma$ introduced in Sec. \ref{sec: gap expansion}.

To make analytical progress, we use the following band dispersions:
\begin{equation}
\label{xi-12-model}
  \xi_1(\bk)=\xi(\bk)-\frac{\Eb}{2},\quad \xi_2(\bk)=\xi(\bk)+\frac{\Eb}{2},
\end{equation}
where $\Eb>0$ is the band splitting, which satisfies 
\begin{equation}
\label{assumptions}
  \Eb<2\epsilon_c\ll\epsilon_F. 
\end{equation}
Under these assumptions, the Cooper pairing, both intraband and interband, takes place within a ``thick'' momentum shell containing both Fermi surfaces, 
as shown in Fig. \ref{fig: Fermi-surfaces}. The BCS cutoffs in Eq. (\ref{V-BCS-cutoffs}) take the same form in both bands and also appear in the basis functions as $\hat\phi_{nn'}^{a}(\bk)\propto\theta(\epsilon_c-|\xi(\bk)|)$.  
The relative strength of the intraband and interband pairing is controlled by the coupling constants. 

The smallness of the superconducting energy scales compared to the Fermi energy allows 
one to transform the momentum integrals that appear in the calculations below as follows:
\begin{equation}
\label{integral-cutoff}
  \int\frac{d^2\bk}{(2\pi)^2}\,(...)=N_F\int_{-\epsilon_c}^{\epsilon_c} d\xi\,\left\langle(...)\right\rangle,
\end{equation}
where $N_F=\int_{\bk}\delta[\xi(\bk)]$ is the density of states (DoS) corresponding to $\xi(\bk)$ and the angular brackets denote the average over the constant-energy surface $\xi(\bk)=0$. It should be noted that only the assumption that both Fermi surfaces are inside the same BCS shell is crucial for our analysis. 
The assumption $\epsilon_c\ll\epsilon_F$ facilitates analytical calculations but can be relaxed, e.g., one can extend the energy cutoff to the bandwidth if needed.

One can use Eqs. (\ref{c-transform-g}) and (\ref{phi-transform-g}) to show that the factorized form (\ref{V-factorized}) of the pairing interaction ensures that the Hamiltonian is invariant under the point group operations: 
$g\hat H_{int}g^{-1}=\hat H_{int}$. The coupling constants satisfy the constraints
\begin{equation}
\label{V1234-Hermiticity}
  V_{n_1n_2,n_3n_4}=V^*_{n_4n_3,n_2n_1},
\end{equation}
which follows from the Hermiticity of $\hat H_{int}$, and also 
\begin{equation}
\label{V1234-anticommutation}
  V_{n_1n_2,n_3n_4}=V_{n_2n_1,n_3n_4}=V_{n_1n_2,n_4n_3},
\end{equation}
which follows from the anticommutation of the fermionic operators. The final set of constraints,
\begin{equation}
\label{V1234-TR}
  V_{n_1n_2,n_3n_4}=V^*_{n_2n_1,n_4n_3},
\end{equation}
comes from the requirement that the Hamiltonian is invariant under TR, i.e., $K\hat H_{int}K^{-1}=\hat H_{int}$, after Eq. (\ref{K-constraint-basis-functions}) is taken into account. 
Combining Eqs. (\ref{V1234-Hermiticity}), (\ref{V1234-anticommutation}), and (\ref{V1234-TR}), we see that the coupling constants for all band combinations are real and have the following symmetry properties:
\begin{equation}
\label{V1234-final-constraints}
  V_{n_1n_2,n_3n_4}=V_{n_2n_1,n_3n_4}=V_{n_1n_2,n_4n_3}=V_{n_3n_4,n_1n_2}.
\end{equation}
Therefore, in the two-band case there are six independent coupling constants: $V_{11,11}$, $V_{22,22}$, $V_{11,22}$, $V_{11,12}$, $V_{12,12}$, and $V_{12,22}$, the last three describing the pairing of quasiparticles 
from different bands. The constants $V_{11,11}$ and $V_{22,22}$ describe the intraband pairing in the bands 1 and 2, respectively, whereas $V_{11,22}$ describes the pair scattering (the Josephson coupling) between different bands.
In Appendix \ref{app: V-example}, we calculate the coupling constants in a simple model of the pairing interaction which is local is real space.

\subsection{Gap equations}
\label{sec: self-consistency}

In the remainder of the paper, we focus on the 1D pairing channels, exemplified by the $s$-wave pairing, see Sec. \ref{sec: s-wave-D-4h}. 
The order parameter has three components, two intraband and one interband, see Eq. (\ref{s-wave-OP-psis}), which can be written in a compact form as 
$$
  \bmeta(\br)=\left(\begin{array}{c}
          \eta_{1}(\br) \\ \eta_{2}(\br) \\ \tilde\eta(\br)
         \end{array}\right),\quad \bmeta(\bq)=\frac{1}{\cal V}\int d^2\br\,\bmeta(\br)e^{-i\bq\br}.
$$
We shall also use the following shorthand notation for the six independent coupling constants:
\begin{eqnarray*}
  & V_{11,11}=V_{11},\quad V_{22,22}=V_{22},\quad V_{11,22}=V_{12},\\
  & V_{11,12}=\tilde V_{13},\quad V_{22,12}=\tilde V_{23},\quad V_{12,12}=\tilde V_{33},
\end{eqnarray*}
which can be combined into a real positive-definite symmetric matrix
\begin{equation}
\label{W-matrix-123}
  \hat W=\left(\begin{array}{ccc}
	V_{11} & V_{12} & \tilde V_{13} \\
	V_{12} & V_{22} & \tilde V_{23} \\
	\tilde V_{13} & \tilde V_{23} & \tilde V_{33} \\
	\end{array}\right).
\end{equation}
To describe the ``usual'' two-band superconductor without the interband pairing, all quantities with a tilde, namely, $\tilde V_{13}$, $\tilde V_{23}$, $\tilde V_{33}$, and $\tilde\eta$, should be set to zero.

In a uniform superconducting state, we have $\bmeta(\bq)=\bmeta\delta_{\bq,0}$ and, according to Appendix \ref{app: effective action}, the self-consistency equations for the gap functions take the following form:  
\begin{equation}
\label{gap-eq-s-wave}
	\hat W^{-1}\bmeta = \frac{1}{2}T\sum_m\int\frac{d^2\bk}{(2\pi)^2}\,\tr\biggl(\frac{\partial\hat{\cal G}^{-1}}{\partial\bmeta^{*}}\hat{\cal G}\biggr),
\end{equation}
where $\omega_m=(2m+1)\pi T$ is the fermionic Matsubara frequency and
\begin{widetext}
\begin{equation}
\label{GF88-s-wave}
  \hat{\cal G}^{-1}(\bk,\omega_m)=\left( \begin{array}{cccc}
	i\omega_m-\xi_1(\bk) & -\hat\Delta_{11}(\bk) & 0 & -\hat\Delta_{12}(\bk) \\
	-\hat\Delta^\dagger_{11}(\bk) & i\omega_m+\xi_1(\bk) & -\hat\Delta^\dagger_{21}(\bk)  & 0 \\
	0 & -\hat\Delta_{21}(\bk) & i\omega_m-\xi_2(\bk) & -\hat\Delta_{22}(\bk) \\
	-\hat\Delta^\dagger_{12}(\bk) & 0 & -\hat\Delta^\dagger_{22}(\bk) & i\omega_m+\xi_2(\bk) 
	\end{array} \right)
\end{equation}
\end{widetext}
is the inverse Green's function. The intraband and interband gap functions are given by Eq. (\ref{two-band-gaps-s-wave}). The momentum integrals is Eq. (\ref{gap-eq-s-wave}) and everywhere below are understood 
to include the cutoff, as in Eq. (\ref{integral-cutoff}).

Since all $2\times 2$ Kramers blocks in the matrix (\ref{GF88-s-wave}) commute with each other, its inverse can be calculated analytically producing a system of three coupled nonlinear equations for the order parameter components:
\begin{equation}
\label{gap-eq-s-wave-final}
	\left( \begin{array}{c}
	\eta_1 \\ \eta_2 \\ \tilde\eta
	\end{array}\right)=
	T\sum_m\int\frac{d^2\bk}{(2\pi)^2}\;\hat W
	\left( \begin{array}{c}
	\Theta_1(\bk,\omega_m) \\ \Theta_2(\bk,\omega_m) \\ 2\tilde\Theta(\bk,\omega_m)
	\end{array}\right),
\end{equation}
where
\begin{eqnarray*}
  & \Theta_1=\dfrac{\eta_1\alpha_1^2(\omega_m^2+\xi_2^2+|\Delta_2|^2)-\tilde\eta^2\eta_2^*\alpha_1\alpha_2g^2}{\omega_m^4+2P\omega_m^2+Q},\\
  & \Theta_2=\dfrac{\eta_2\alpha_2^2(\omega_m^2+\xi_1^2+|\Delta_1|^2)-\tilde\eta^2\eta_1^*\alpha_1\alpha_2g^2}{\omega_m^4+2P\omega_m^2+Q},\\
  & \tilde\Theta=\dfrac{\tilde\eta g^2(\omega_m^2+\xi_1\xi_2+|\tilde\Delta|^2)
      -\eta_1\eta_2\tilde\eta^*\alpha_1\alpha_2g^2}{\omega_m^4+2P\omega_m^2+Q}.
\end{eqnarray*}
Other notations are as follows: 
\begin{eqnarray*}
    & \Delta_n(\bk)=\eta_n\alpha_n(\bk),\\
    & \tilde\Delta(\bk)=\tilde\eta g(\bk),\quad g=\sqrt{\tilde\alpha^2+\tilde{\bm{\beta}}^2},\\
    & P=\dfrac{1}{2}\left(\xi_1^2+|\Delta_1|^2+\xi_2^2+|\Delta_2|^2\right)+|\tilde\Delta|^2,\\ 
    & Q=r_1^2+r_2^2+r_3^2,
\end{eqnarray*}
and 
\begin{eqnarray}
\label{r_123}
  & r_1=\xi_1\xi_2-|\Delta_1\Delta_2|+\tilde\Delta^2,\smallskip\nonumber \\ 
  & r_2=\xi_1|\Delta_2|+\xi_2|\Delta_1|,\smallskip\\
  & r_3=\sqrt{2g^2\left[|\Delta_1\Delta_2||\tilde\eta|^2-\re(\Delta_1\Delta_2\tilde\eta^{*,2})\right]}.\nonumber
\end{eqnarray}

As a side note, the inverse Green's function (\ref{GF88-s-wave}) can also be used to obtain the energies $E(\bk)$ of the Bogoliubov quasiparticles in the bulk, 
by solving the equation $\det\hat{\cal G}^{-1}(\bk,\omega_m)|_{i\omega_m\to E+i0}=0$. In this way, we find that the Bogoliubov spectrum consists of four twofold degenerate branches $\pm E_{\pm}$, where
\begin{equation}
\label{Bogoliubov-branches}
  E_{\pm}(\bk) = \sqrt{P(\bk)\pm\sqrt{P^2(\bk)-Q(\bk)}}.
\end{equation}
The upper Bogoliubov excitation branch $E_+$ is fully gapped in the superconducting state, but the lower branch $E_-$ vanishes at the wave vector $\bk$ if 
$r_1(\bk) = r_2(\bk) = r_3(\bk) = 0$, corresponding to a gap node. For a detailed investigation of the nodal structure of superconductors with the interband pairing, see Refs. \onlinecite{Sam20} and \onlinecite{Holst23}. 

In the general case, i.e., when the intraband and interband coupling constants are present in the matrix $\hat W$, all three components of $\bmeta$ are nonzero. 
In the limit of purely intraband pairing, we have $\tilde V_{13}=\tilde V_{23}=\tilde V_{33}=0$ and $\tilde\eta=0$, so that the Eq. (\ref{gap-eq-s-wave-final}) is reduced to the following form:
\begin{equation}
\label{intraband-only-gap-eqs}
    \left( \begin{array}{c}
	\eta_1 \\ \eta_2
	\end{array}\right)=
    \left(\begin{array}{cc}
	V_{11} & V_{12} \\
	V_{12} & V_{22}
	\end{array}\right)
	\left( \begin{array}{c}
	{\cal I}_1 \\ {\cal I}_2
	\end{array}\right),
\end{equation}
where
$$
  {\cal I}_n=\frac{1}{2}\eta_n\int\frac{d^2\bk}{(2\pi)^2}\,\alpha_n^2\,\frac{\tanh(\sqrt{\xi_n^2+|\Delta_n|^2}/2T)}{\sqrt{\xi_n^2+|\Delta_n|^2}}.
$$
These are the standard gap equations for a two-band superconductor with the intraband pairing and the interband Josephson coupling characterized by $V_{12}$ (Ref. \onlinecite{Suhl59}). The phase transition is of the second order 
and the critical temperature $T_c$ is found from the linearized version of Eq. (\ref{intraband-only-gap-eqs}).

The gap equations (\ref{gap-eq-s-wave-final}) remain invariant under a simultaneous rotation of the phases of the order parameter components $\eta_1$, $\eta_2$, and $\tilde\eta$ by the same amount. Therefore, 
$\tilde\eta$ can be set to be real positive, but $\eta_1$ and $\eta_2$ can be complex:
\begin{equation}
\label{etas-phases}
  \eta_1=|\eta_1|e^{i\varphi_1},\quad \eta_2=|\eta_1|e^{i\varphi_2}. 
\end{equation}
In a TR-invariant superconducting state, all three components are real and $\varphi_{1,2}=0$ or $\pi$. Stable states that break TR symmetry are also possible, see Sec. \ref{sec: stable states}. 

Due to its complexity, the system of the nonlinear gap equations (\ref{gap-eq-s-wave-final}) with all the coupling constants present is not the most convenient starting point for studying the physics of our superconductor.
For this reason, below we use the Ginzburg-Landau formalism, i.e., assume that the phase transition is of the second order and that the free energy in the vicinity of the critical temperature $T_c$ can be expanded 
in powers of the order parameter $\bmeta(\br)$ and its gradients.

\section{Ginzburg-Landau free energy}
\label{sec: GL functional}

We focus on a two-band $s$-wave superconductor, with the intraband and interband gap functions discussed in Sec. \ref{sec: s-wave-D-4h}. Generalization to other pairing channels is straightforward.
The GL free energy density $F_{GL}=F_2+F_4$ near the superconducting phase transition can be derived using the effective action formalism. Since the technical steps are more or less standard, they are 
relegated to Appendix \ref{app: effective action}. 

The terms quadratic in the order parameter have the form
\begin{equation}
\label{F-GL-quadratic}
  F_2=\bmeta^\dagger\hat A\bmeta+K_1|{\bm\nabla}\eta_1|^2+K_2|{\bm\nabla}\eta_2|^2+\tilde K|{\bm\nabla}\tilde\eta|^2.
\end{equation}
The temperature dependence enters only the uniform terms through
\begin{equation}
\label{A-matrix}
  \frac{1}{N_F}\hat A(T)=\hat\Lambda^{-1}-\left(\begin{array}{ccc}
                                   L(T) & 0 & 0 \\
                                   0 & L(T) & 0 \\
                                   0 & 0 & 2\tilde L(T)
                                   \end{array}\right),
\end{equation}
where 
\begin{equation}
\label{Lambda-def}
  \hat\Lambda=N_F\hat W=\left(\begin{array}{ccc}
	\lambda_{11} & \lambda_{12} & \tilde\lambda_{13} \\
	\lambda_{12} & \lambda_{22} & \tilde\lambda_{23} \\
	\tilde\lambda_{13} & \tilde\lambda_{23} & \tilde\lambda_{33} \\
	\end{array}\right)
\end{equation}
is a symmetric matrix of the dimensionless coupling constants, with $\hat W$ given by Eq. (\ref{W-matrix-123}), and
\begin{equation}
\label{L-tilde L-defs}
  \left.\begin{array}{l}
        L(T)=\ln\left(\dfrac{2e^{\mathbb{C}}\epsilon_c}{\pi T}\right),\smallskip\\
        \tilde L(T)=\ln\left(\dfrac{2e^{\mathbb{C}}\epsilon_c}{\pi T}\right)+\Psi\left(\frac{1}{2}\right)-\re\Psi\left(\dfrac{1}{2}-i\dfrac{\Eb}{4\pi T}\right),
  \end{array}\right.
\end{equation}
where $\mathbb{C}\simeq 0.577$ is Euler's constant and $\Psi(z)$ is the digamma function. We assume that the intraband and interband basis functions are normalized as follows: 
$\langle\alpha_1^2(\bk)\rangle=\langle\alpha_2^2(\bk)\rangle=\langle g^2(\bk)\rangle=1$. Note that $L$ diverges at $T\to 0$, which corresponds to the standard logarithmic singularity in the intraband Cooper channel, whereas $\tilde L$ does not diverge, since the singularity in the interband Cooper channel is cut off by the band splitting. The gradient terms in Eq. (\ref{F-GL-quadratic}) are discussed in Sec. \ref{sec: gradient terms} below. 

The expressions (\ref{F-GL-quadratic}) and (\ref{A-matrix}) are valid only if the coupling constants form an invertible matrix. In the purely intraband limit, we have
$\tilde\lambda_{13}=\tilde\lambda_{23}=\tilde\lambda_{33}=0$ and the matrix $\hat\Lambda$ is singular. In this case, $\tilde\eta$ identically vanishes and the uniform terms in Eq. (\ref{F-GL-quadratic}) take the from
\begin{equation}
\label{F-GL-quadratic-intraband}
  F_2=\alpha_1(T)|\eta_1|^2+\alpha_2(T)|\eta_2|^2+\gamma(\eta_1^*\eta_2+\mathrm{c.c.}),
\end{equation}
where
\begin{eqnarray*}
  && \alpha_1(T)=\left[\frac{\lambda_{22}}{\lambda_{11}\lambda_{22}-\lambda_{12}^2}-L(T)\right]N_F,\\
  && \alpha_2(T)=\left[\frac{\lambda_{11}}{\lambda_{11}\lambda_{22}-\lambda_{12}^2}-L(T)\right]N_F,\\
  && \gamma=-\frac{\lambda_{12}}{\lambda_{11}\lambda_{22}-\lambda_{12}^2}N_F.
\end{eqnarray*}
Thus, the usual GL theory for a two-band superconductor with the Josephson coupling between the bands is recovered.\cite{Tilley64,GZK67,Zh04} 
The opposite case of a purely interband pairing (only $\tilde\lambda_{33}\neq 0$) is discussed in Sec. \ref{sec: interband-dominant}.

\subsection{Critical temperature}
\label{sec: Tc}

In general, all six coupling constants in the matrix (\ref{Lambda-def}) are nonzero. Phenomenologically, they are constrained only by the requirement that $\hat\Lambda$ is real and positive-definite, which means, in particular, 
that the diagonal elements are all positive. The off-diagonal elements can have either sign. 

At sufficiently high temperatures, the matrix $\hat A$ is positive-definite and the minimum of the free energy is achieved at $\bmeta=\bm{0}$, i.e., in the normal state.
As the temperature is lowered, one of the eigenvalues of $\hat A$ changes sign, so that the critical temperature $T_c$ of the second-order phase transition into a uniform superconducting state is found by solving the equation
\begin{equation}
\label{Tc-equation}
  \det\hat A(T)=0.
\end{equation}
It can be shown that the maximum critical temperature corresponds to the state in which all three components of the order parameter are nonzero, 
see Appendix \ref{app: maximum-Tc}. In other words, if all coupling constants are nonzero, then the normal-state instability towards the general superconducting state $(\eta_1,\eta_2,\tilde\eta)$ 
occurs at a higher temperature than the instability towards a reduced state $(\eta_1,\eta_2,0)$. This can be interpreted as an ``enhancement'' of superconductivity by the interband pairing.

At given coupling constants, the onset of superconductivity is controlled by the interband splitting $\Eb$. We observe that 
$$
  \frac{dT_c}{d\Eb}=-\frac{\partial(\det\hat A)/\partial\Eb}{\partial(\det\hat A)/\partial T}\biggr|_{T=T_c}.
$$
The denominator here is positive, while for the numerator we obtain from Eqs. (\ref{A-matrix}) and (\ref{L-tilde L-defs}):
\begin{eqnarray*}
  && \frac{\partial}{\partial\Eb}\det\hat A = \tr\biggl(\mathrm{adj\,}\hat A\,\frac{\partial\hat A}{\partial\Eb}\biggr)\\
  && \qquad=\frac{1}{2\pi T}(A_{11}A_{22}-A_{12}^2)\im\Psi'\left(\frac{1}{2}-i\frac{\Eb}{4\pi T}\right),
\end{eqnarray*}
where $\mathrm{adj\,}\hat A$ is the adjugate matrix. The function $\im\Psi(1/2-ix)$ is positive at all $x>0$ and it follows from Sylvester's criterion that the principal minors of $\hat A$, including 
$A_{11}A_{22}-A_{12}^2$, are positive at $T_c$. Therefore,
\begin{equation}
\label{Tc-suppressed-by-Eb}
  \frac{dT_c}{d\Eb}<0,
\end{equation}
i.e., increasing the band splitting $\Eb$ always suppresses the critical temperature, regardless of all other parameters of the system, including the relative magnitudes of the intraband and interband coupling constants. 

After some straightforward manipulations, Eq. (\ref{Tc-equation}) takes the form 
\begin{equation}
\label{Tc-equation-1}
  AL^2-2BL+1=0,
\end{equation}
where
\begin{eqnarray*}
  A &=& \lambda_{11}\lambda_{22}-\lambda_{12}^2 \\
  && +\frac{2(\lambda_{11}\tilde\lambda_{23}^2+\lambda_{22}\tilde\lambda_{13}^2-2\lambda_{12}\tilde\lambda_{13}\tilde\lambda_{23})\tilde L}{1-2\tilde\lambda_{33}\tilde L},\\
  B &=& \frac{\lambda_{11}+\lambda_{22}}{2}+\frac{(\tilde\lambda_{13}^2+\tilde\lambda_{23}^2)\tilde L}{1-2\tilde\lambda_{33}\tilde L}.
\end{eqnarray*}
A closed-form expression for the critical temperature can be obtained only in the limit $T_c\ll\Eb$. Using the asymptotic form $\Psi(z)\simeq\ln z$ at $z\to\infty$ (Ref. \onlinecite{AS65}), 
we find that $\tilde L$ attains a finite temperature-independent value:
\begin{equation}
\label{tilde-L-T-independent}
  \tilde L_0\equiv\tilde L(0)=\ln\frac{2\epsilon_c}{\Eb}\geq 0.
\end{equation}
Now the equation (\ref{Tc-equation-1}) can be easily solved, with the following result:
\begin{equation}
\label{T_c-general}
  T_c=\frac{2e^{\mathbb{C}}\epsilon_c}{\pi}e^{-1/\lambda},\quad \lambda=B+\sqrt{B^2-A}.
\end{equation}
We assume that all six dimensionless coupling constants in Eq. (\ref{Lambda-def}) are small in magnitude and that $\lambda\ll 1$, which corresponds to the weak-coupling limit, in which $T_c\ll\Eb<2\epsilon_c$. 

The effects of the interband pairing, which are contained in the last terms in $A$ and $B$, depend on the six coupling constants and also on the band splitting $\Eb$, making it difficult to characterize them by 
a simple single parameter.  
To make progress, we assume that either all six coupling constants have the same order of magnitude (which they do if the pairing is local in real space, see Appendix \ref{app: V-example}), or 
the three ``interband'' constants differ from the three ``intraband'' ones by a factor 
\begin{equation}
\label{tilde-omega}
  \tilde\omega=\frac{\max(|\tilde\lambda_{13}|,|\tilde\lambda_{23}|,\tilde\lambda_{33})}{\max(\lambda_{11},\lambda_{22},|\lambda_{12}|)}.
\end{equation}
Another dimensionless parameter that appears in $A$ and $B$,
\begin{equation}
\label{tilde-rho}
  \tilde\rho=\max(|\tilde\lambda_{13}|,|\tilde\lambda_{23}|,\tilde\lambda_{33})\,\tilde L_0,
\end{equation}
can be used as a measure of the effect of the band splitting. It is reasonable to assume that $\tilde\rho\ll 1$ in the weak-coupling theory, the assumption that would break down only if the band splitting 
is exponentially small compared to $\epsilon_c$.

\subsubsection{Dominant intraband pairing}
\label{sec: intraband-dominant}

If the last terms in $A$ and $B$ represent small corrections then the critical temperature is largely determined by the intraband coupling constants. In terms of the parameters (\ref{tilde-omega}) and (\ref{tilde-rho}), this corresponds to $\tilde\omega\tilde\rho\ll 1$, in which case
\begin{equation}
\label{g-dominant-intraband}
  \lambda=\lambda_0[1+{\cal O}(\tilde\omega\tilde\rho)],
\end{equation}
where 
$$
  \lambda_0=\frac{\lambda_{11}+\lambda_{22}}{2}+\sqrt{\left(\frac{\lambda_{11}-\lambda_{22}}{2}\right)^2+\lambda_{12}^2}
$$
is the effective coupling constant in a two-band superconductor without interband pairing. If $\tilde\omega\ll 1$, then $T_c\to T_{c0}+0$, where $T_{c0}=(2e^{\mathbb{C}}\epsilon_c/\pi)e^{-1/\lambda_0}$ is the ``intraband-only'' critical temperature.

At fixed coupling constants, the critical temperature of the three-component superconducting state $\bmeta=(\eta_1,\eta_2,\tilde\eta)$ is suppressed by increasing the band splitting, until at $\Eb=2\epsilon_c$ we have
$\tilde L_0=0$, the interband component disappears altogether, and the phase transition takes place at $T_{c0}$ into the reduced state with $\bmeta=(\eta_1,\eta_2,0)$. 
Further increasing $\Eb$ does not affect the critical temperature, because the pairing shells in the two bands no longer overlap.

\subsubsection{Dominant interband pairing}
\label{sec: interband-dominant}

The extreme limit of a purely interband pairing is realized when the only nonzero coupling constant is $\tilde\lambda_{33}$, so that $\tilde\omega\to\infty$. Then, the order parameter has only one component $\tilde\eta$ and the critical temperature equation (\ref{Tc-equation-1}) takes the following form:
$$
  \tilde L(T)=\frac{1}{2\tilde\lambda_{33}}.
$$
Since $\max_T\tilde L(T)=\tilde L_0+{\cal O}(1)$, the last equation does not have a solution if 
$\tilde\lambda_{33}\tilde L_0\ll 1$. Therefore, the purely interband superconductivity is completely suppressed by a sufficiently large band splitting, which is is analogous to the paramagnetic pair breaking by a strong 
magnetic field in the usual BCS case,\cite{Sarma63,Tinkham-book} with $\Eb$ playing the role of the Zeeman magnetic field.  
At $\tilde\lambda_{33}\tilde L_0\ll 1$, a nonuniform interband superconductivity of the FFLO type\cite{FF64,LO64} is also suppressed.

Let us now suppose that the intraband coupling constants are nonzero but small compared with the interband ones, so that $\tilde\omega\tilde\rho\gg 1$. In this case, all three components of the order parameter are nonzero and the critical temperature is given by Eq. (\ref{T_c-general}), with the effective coupling constant
\begin{equation}
\label{g-dominant-interband}
  \lambda=2(\tilde\lambda_{13}^2+\tilde\lambda_{23}^2)\tilde L_0.
\end{equation}
If $\tilde\lambda_{13}=\tilde\lambda_{23}=0$, then $\lambda$ and therefore $T_c$ vanish, in agreement with the complete suppression of superconductivity in the purely interband limit.

\subsection{Gradient terms}
\label{sec: gradient terms}

The coefficients in the gradient energy can be evaluated at the critical temperature, with the following result, see Appendix \ref{app: effective action} for details:
\begin{equation}
\label{K-coefficients}
  \left.\begin{array}{c}
	K_1=\dfrac{7\zeta(3)N_Fw_1}{16\pi^2T_c^2},\quad K_2=\dfrac{7\zeta(3)N_Fw_2}{16\pi^2T_c^2},\medskip\\ 
	\tilde K=\dfrac{7\zeta(3)N_F\tilde w}{8\pi^2T_c^2}f_1\left(\dfrac{\Eb}{4\pi T_c}\right),
  \end{array}\right.
\end{equation}
where $\zeta(3)\simeq 1.20$ is the Riemann zeta function, $w_n=\langle\alpha_n^2v_x^2\rangle=\langle\alpha_n^2v_y^2\rangle$, $\tilde w=\langle g^2v_x^2\rangle=\langle g^2v_y^2\rangle$,
and
\begin{equation}
\label{f_1}
  f_1(x) = -\frac{1}{14\zeta(3)}\re\Psi''\left(\frac{1}{2}-ix\right).
\end{equation}
This function is plotted in Fig. \ref{fig: f123}. In particular, in the limit of a large band splitting, $\Eb\gg T_c$, using the asymptotics $f_1(x)\propto-1/x^2$ at $x\gg 1$ (Ref. \onlinecite{AS65}), we find that the coefficient $\tilde K$ 
is much smaller than $K_1$ and $K_2$:
\begin{equation}
\label{tilde K-small}
  \frac{|\tilde K|}{K_{1,2}}\sim\frac{T_c^2}{\Eb^2}.
\end{equation}
Note that $\tilde K$ becomes negative at a sufficiently large band splitting (changing sign at $\Eb/T_c\simeq 3.82$, see Fig. \ref{fig: f123}), 
which indicates a possible instability towards a nonuniform superconducting state even at zero external magnetic field. Let us see if such an instability indeed takes place.

It follows from Eq. (\ref{F-GL-quadratic}) that the critical temperature of a continuous phase transition into a nonuniform state is found from the equation $\det\hat{\cal A}(T,\bq)=0$, where
\begin{equation}
\label{A-T-q}
  \hat{\cal A}(T,\bq)=\hat A(T)+ 
    \left( \begin{array}{ccc}
    K_1q^2 & 0 & 0 \\
    0 & K_2q^2 & 0 \\
    0 & 0 & \tilde Kq^2
    \end{array} \right),  
\end{equation}
where $\hat A$ is given by Eq. (\ref{A-matrix}). Therefore,
\begin{equation}
\label{dTc-dq}
  \frac{dT_c}{dx}\biggl|_{q=0}=-\frac{\partial(\det\hat{\cal A})/\partial x}{\partial(\det\hat{\cal A})/\partial T}\biggl|_{q=0},\quad x=q^2.
\end{equation}
The denominator here is positive, because the matrix $\hat A$ is positive-definite at temperatures above $T_c$, while for the numerator we have
\begin{eqnarray}
\label{derivative-det A}
  && \frac{\partial(\det\hat{\cal A})}{\partial x}\biggl|_{q=0} = (A_{11}A_{22}-A_{12}^2)\tilde K\nonumber\\
  && \quad+(A_{22}A_{33}-A_{23}^2)K_1+(A_{11}A_{33}-A_{13}^2)K_2,\qquad
\end{eqnarray}
where the matrix elements of $\hat A$ are taken at $T=T_c$. 

A nonuniform superconducting state has a higher critical temperature than the uniform one if the derivative (\ref{dTc-dq}) is positive. Since the coefficients multiplying $K_1$, $K_2$, and $\tilde K$ on the right-hand side of 
Eq. (\ref{derivative-det A}) are nothing but the principal minors of $\hat A$, which are all positive at $T_c$, the uniform superconducting state is unstable near $T_c$ if $\tilde K<0$ and 
$$
  |\tilde K|>\frac{K_1(A_{22}A_{33}-A_{23}^2)+K_2(A_{11}A_{33}-A_{13}^2)}{A_{11}A_{22}-A_{12}^2}.
$$
In view of Eq. (\ref{tilde K-small}), this last condition is difficult to satify for a large band splitting. Although one cannot rule out that the nonuniform instability may be present in some fine-tuned ranges of the parameters, 
we shall not further investigate this possibility here.

\subsection{Quartic terms}
\label{sec: GL quartic terms}

\begin{figure}
\includegraphics[width=7cm]{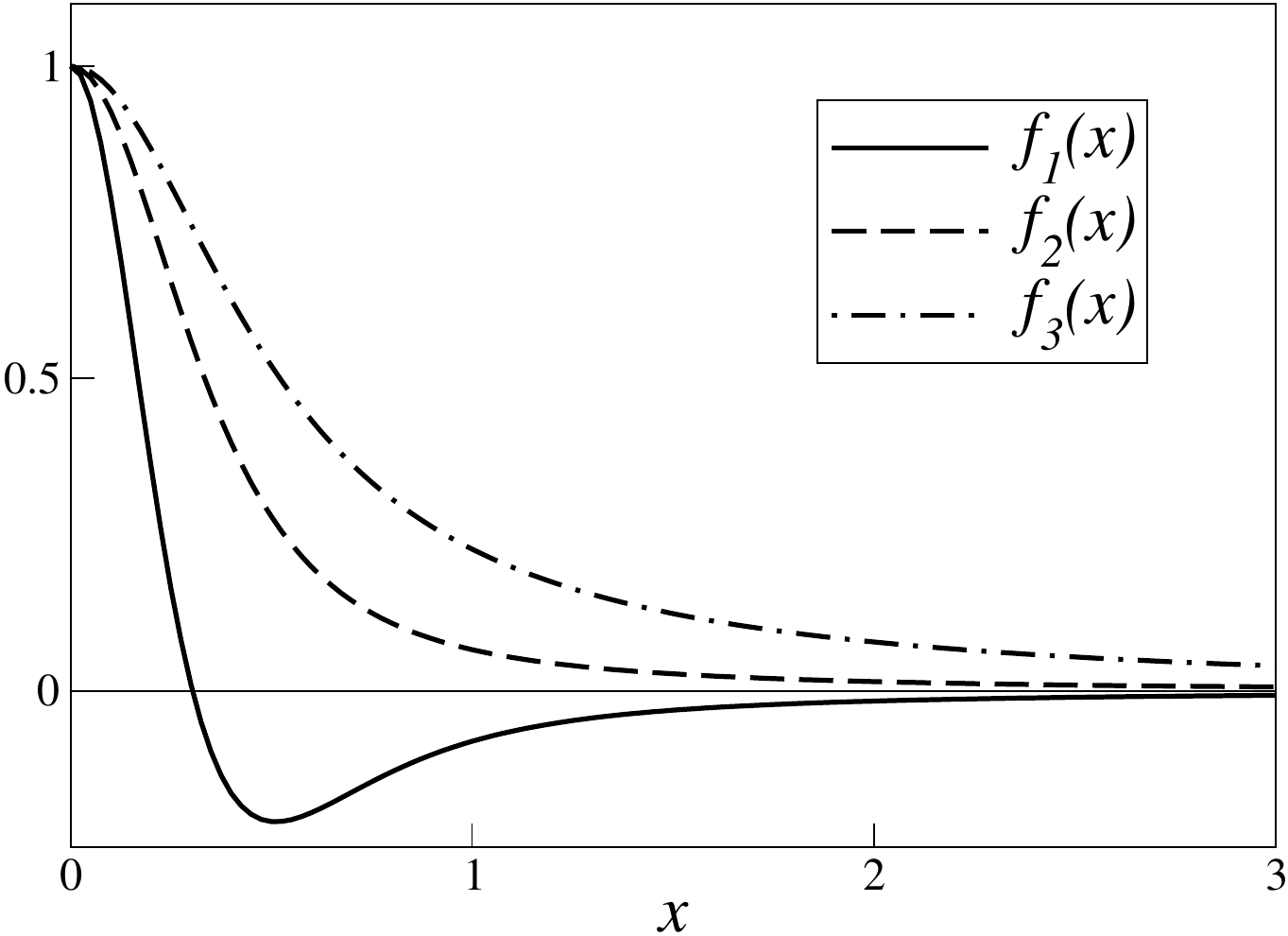}
\caption{The functions $f_{1,2,3}(x)$ which determine the dependence of the GL free energy coefficients on the band splitting $\Eb$, see Eqs. (\ref{K-coefficients}) and (\ref{beta-coefficients}), with $x=\Eb/4\pi T_c$. 
Note that $f_1(x)$ changes sign at $x\simeq 0.30$.}
\label{fig: f123}
\end{figure}

According to Appendix \ref{app: effective action}, the uniform fourth-order terms in the free energy density have the following form:
\begin{eqnarray}
\label{F-GL-quartic-final}
  F_4 = \beta_1|\eta_1|^4+\beta_2|\eta_2|^4+\tilde\beta_1|\eta_1|^2|\tilde\eta|^2+\tilde\beta_2|\eta_2|^2|\tilde\eta|^2 \nonumber\\
  +\tilde\beta_3|\tilde\eta|^4+\tilde\beta_4(\eta_1\eta_2\tilde\eta^{*,2}+\mathrm{c.c.}),
\end{eqnarray}
with
\begin{eqnarray}
\label{beta-coefficients}
  && \beta_1=\beta_0\langle\alpha_1^4\rangle,\  \beta_2=\beta_0\langle\alpha_2^4\rangle, \nonumber\\  
  && \tilde\beta_1=4\beta_0 f_2\left(\frac{\Eb}{4\pi T_c}\right)\langle\alpha_1^2g^2\rangle, \nonumber\\ 
  && \tilde\beta_2=4\beta_0 f_2\left(\frac{\Eb}{4\pi T_c}\right)\langle\alpha_2^2g^2\rangle, \\ 
  && \tilde\beta_3=2\beta_0 f_1\left(\frac{\Eb}{4\pi T_c}\right)\langle g^4\rangle, \nonumber\\
  && \tilde\beta_4=2\beta_0 f_3\left(\frac{\Eb}{4\pi T_c}\right)\langle\alpha_1\alpha_2g^2\rangle. \nonumber
\end{eqnarray}
Here
$$
  \beta_0=\frac{7\zeta(3)N_F}{16\pi^2T_c^2},
$$
$f_1(x)$ is given by Eq. (\ref{f_1}), and the functions
\begin{equation}
\label{f_2}
  f_2(x)=\frac{1}{14\zeta(3)x}\im\Psi'\left(\frac{1}{2}-ix\right)
\end{equation}
and 
\begin{equation}
\label{f_3}
  f_3(x)=\frac{1}{7\zeta(3)x^2}\left[\re\Psi\left(\frac{1}{2}-ix\right)-\Psi\left(\frac{1}{2}\right)\right]
\end{equation}
are plotted in Fig. \ref{fig: f123}.

Generically, $\beta_1$ and $\beta_2$ are of the same order of magnitude: $\beta_1\sim\beta_2\sim\beta_0$. In the limit of a large band splitting ($\Eb\gg T_c$), we use the asymptotics $f_2(x)\propto 1/x^2$ and
$f_3(x)\propto\ln x/x^2$ (Ref. \onlinecite{AS65}) and find that the coefficients in the terms involving the interband pairing are smaller than $\beta_{1,2}$:
$$
  \frac{\tilde\beta_{1,2}}{\beta_0}\sim\frac{T_c^2}{\Eb^2},\quad 
  \frac{|\tilde\beta_3|}{\beta_0}\sim\frac{T_c^2}{\Eb^2},\quad 
  \frac{\tilde\beta_4}{\beta_0}\sim\frac{T_c^2}{\Eb^2}\ln\frac{\Eb}{T_c}.
$$
Note that, similarly to $\tilde K$, the coefficient $\tilde\beta_3$ changes sign at $\Eb/T_c\simeq 3.82$, so that one could ask whether a first-order transition into the interband-only state $\bmeta=(0,0,\tilde\eta)$ can preempt the second-order transition into the general state in which all three components of the order parameter are nonzero. It is easy to see that the answer is negative, since the interband-only state does not satisfy the gap equations (\ref{gap-eq-s-wave-final}) if all coupling constants are present. In the purely interband limit, in which only $\tilde V_{33}$ is nonzero, superconductivity is completely suppressed by the large band splitting, see Sec. \ref{sec: interband-dominant}.

\subsection{Discussion}
\label{sec: GL-summary}

Putting our results together, the GL free energy of a two-band superconductor with interband pairing has the following form:
\begin{equation}
\label{F-GL-FINAL}
  F_{GL}=F_2+F_4,
\end{equation}
where
\begin{eqnarray}
\label{F-GL-2-final}
  F_2 &=& \alpha_1|\eta_1|^2+\alpha_2|\eta_2|^2+\gamma(\eta_1^*\eta_2+\mathrm{c.c.}) \nonumber\\
    && +\tilde\alpha|\tilde\eta|^2+\tilde\gamma_1(\eta_1^*\tilde\eta+\mathrm{c.c.})+\tilde\gamma_2(\eta_2^*\tilde\eta+\mathrm{c.c.}) \nonumber\\
    && +K_1|{\bm\nabla}\eta_1|^2+K_2|{\bm\nabla}\eta_2|^2+\tilde K|{\bm\nabla}\tilde\eta|^2
\end{eqnarray}
and $F_4$ is given by Eq. (\ref{F-GL-quartic-final}). The coefficients in the uniform quadratic terms are given by
\begin{eqnarray*}
  & \alpha_n=[(\hat\Lambda^{-1})_{nn}-L(T)]N_F,\\
  & \tilde\alpha=[(\hat\Lambda^{-1})_{33}-2\tilde L(T)]N_F,\\
  & \gamma=(\hat\Lambda^{-1})_{12}N_F,\quad \tilde\gamma_n=(\hat\Lambda^{-1})_{n3}N_F.
\end{eqnarray*}
We see that, while $\alpha_n$ changes sign at the temperature $T_n=(2e^{\mathbb{C}}\epsilon_c/\pi)\exp[-(\hat\Lambda^{-1})_{nn}]$, the temperature dependence of $\tilde\alpha$ is negligible if $\Eb\gg T_c$, when $\tilde L(T)\simeq\tilde L_0$, see Eq. (\ref{tilde-L-T-independent}). Therefore, in the large band splitting limit, $\tilde\alpha$ is just a positive constant, which is consistent with the fact that superconductivity is completely suppressed in the interband-only case, see Sec. \ref{sec: interband-dominant}.

From the symmetry point of view, the phenomenological GL free energy can contain many more terms than those listed in Eqs. (\ref{F-GL-2-final}) and (\ref{F-GL-quartic-final}). 
Namely, any combination of the order parameter components and their gradients which is (i) real, (ii) invariant under all operations of the point group, and (iii) invariant under a simultaneous rotation of 
the phases of $\eta_1$, $\eta_2$, and $\tilde\eta$, can appear in $F_{GL}$. For example, such quartic terms as
$$
  \eta_1^2\eta_2^{*,2}+\mathrm{c.c.},\ \eta_n^2\tilde\eta^{*,2}+\mathrm{c.c.},\ |\tilde\eta|^2(\eta_1\eta_2^*+\mathrm{c.c.}),
$$
are all allowed by symmetry, as are the gradient terms 
$$
  ({\bm\nabla}\eta_1)^*({\bm\nabla}\eta_2)+\mathrm{c.c.},\ ({\bm\nabla}\eta_n)^*({\bm\nabla}\tilde\eta)+\mathrm{c.c.}. 
$$
Our microscopic derivation shows that, in order to obtain any of these additional terms, one has to go beyond the extended BCS framework. 

A two-band superconductor with interband pairing is characterized by a three-component order parameter $\bmeta=(\eta_1,\eta_2,\tilde\eta)$, so it is natural to ask 
how our results compare with the GL energy for a three-band superconductor \textit{without} interband pairing.
In the latter case, the order parameter also has three components, $\eta_1$, $\eta_2$, and $\eta_3$, which describe the intraband pair condensates in each of the bands, and the free energy density is given by a straightforward extension of the standard two-band GL theory:
\begin{eqnarray}
\label{F-GL-3-band}
  F^{\mathrm{3-band}}_{GL}=\sum_{n=1}^3\left[\alpha_n|\eta_n|^2+K_n|{\bm\nabla}\eta_n|^2+\beta_n|\eta_n|^4\right] \nonumber\\
  +\sum_{n\neq n'} \gamma_{nn'}(\eta_n^*\eta_{n'}+\mathrm{c.c.}),
\end{eqnarray}
see, e.g., Refs. \onlinecite{Lin14}, \onlinecite{Tanaka15}, and \onlinecite{ABG99}.
The differences between Eqs. (\ref{F-GL-FINAL}) and (\ref{F-GL-3-band}) can summarized as follows. In Eq. (\ref{F-GL-2-final}), only the intraband coefficients $\alpha_1$ and $\alpha_2$ essentially depend on temperature, 
but not $\tilde\alpha$, in contrast to Eq. (\ref{F-GL-3-band}), in which all three coefficients $\alpha_{1,2,3}$ are $T$-dependent. In the gradient terms in 
Eq. (\ref{F-GL-2-final}), the coefficient $\tilde K$ becomes negative and much smaller than $K_1$ and $K_2$ in the large band splitting limit, whereas all three gradient terms in $F^{\mathrm{3-band}}_{GL}$ 
are positive and generally comparable in magnitude. Finally, the fourth-order terms (\ref{F-GL-quartic-final}) have a more complicated structure than those in Eq. (\ref{F-GL-3-band}). 
This leads to a rich variety of stable superconducting states, including those that break TR symmetry, see the next section.

\section{TR symmetry-breaking states}
\label{sec: stable states}

The GL free energy expansion (\ref{F-GL-FINAL}) is quantitatively valid in the vicinity of the critical temperature $T_c$, where it can be used to show that $\bmeta\propto(T_c-T)^{1/2}$, with all three components real, see Appendix \ref{app: OP-near-Tc}. In this section, we assume that the physics of our superconductor at temperatures much lower than $T_c$ can also be captured, at least qualitatively, by the free energy (\ref{F-GL-FINAL}), which is treated in the London approximation. Namely, we fix the order parameter magnitudes and minimize $F_{GL}$ only with respect to the relative phases of the order parameter components. 

Setting $\tilde\eta$ to be real positive and using Eq. (\ref{etas-phases}), the uniform phase-dependent terms in the free energy density take the following form:
\begin{eqnarray}
\label{F-phi1-phi2}
  F_{phase} = J\cos(\varphi_1-\varphi_2)+\tilde J_1\cos\varphi_1+\tilde J_2\cos\varphi_2 \nonumber\\
  +\tilde J_3\cos(\varphi_1+\varphi_2),
\end{eqnarray}
where 
\begin{eqnarray*}
  & J=2\gamma|\eta_1||\eta_2|,\quad \tilde J_1=2\tilde\gamma_1|\eta_1|\tilde\eta,\quad \tilde J_2=2\tilde\gamma_2|\eta_2|\tilde\eta,\\
  & \tilde J_3=2\tilde\beta_4|\eta_1||\eta_2|\tilde\eta^2.
\end{eqnarray*}
The first term in Eq. (\ref{F-phi1-phi2}) has the form usual for two-band superconductors, with $J$ corresponding to the interband Josephson coupling, whereas the remaining terms describe the effects of the interband pairing.
If the latter is neglected, then there are only two uniform stable states: $\varphi_1-\varphi_2=\pi$ for $J>0$ and $\varphi_1-\varphi_2=0$ for $J<0$, 
both of which preserve TR symmetry. It follows from the results of the previous section that $\tilde J_3$ is positive, whereas $J$, $\tilde J_1$, and $\tilde J_2$ can have either sign. 

Before we proceed with finding the stable minima of the free energy (\ref{F-phi1-phi2}), we note that TR symmetry-breaking states have been extensively studied in three-band superconductors with only intraband pairing.\cite{ABG99,ST10,TY10}
The three-band London energy obtained from Eq. (\ref{F-GL-3-band}) depends on the condensate phases $\varphi_1$ and $\varphi_2$ in two of the bands (one can set the phase in the third band to zero) and looks similar
to Eq. (\ref{F-phi1-phi2}), but with $\tilde J_3=0$. The TR symmetry-breaking states in the three-band model can be realized when the order parameter phases are ``frustrated'', i.e., when $\sign(J\tilde J_1\tilde J_2)>0$. 

To make analytical progress, we assume that the intraband parameters are the same in both bands, so that $\tilde J_1=\tilde J_2$. Writing $F_{phase}=|J|f(\varphi_1,\varphi_2)$, we have to minimize the following 
function:
\begin{eqnarray}
\label{f-phase-pm}
  f(\varphi_1,\varphi_2)=\sigma\cos(\varphi_1-\varphi_2)+p(\cos\varphi_1+\cos\varphi_2)\nonumber\\
  +q\cos(\varphi_1+\varphi_2),
\end{eqnarray}
where $\sigma=\sign(J)$ and
$$
  p=\frac{\tilde J_1}{|J|}=\frac{\tilde J_2}{|J|},\quad q=\frac{\tilde J_3}{|J|}>0.
$$
If $(\varphi_1,\varphi_2)$ is a critical point of $f$, then $(\varphi_1+\pi,\varphi_2+\pi)$ is a critical point of $f$ with $p$ replaced by $-p$. Therefore, when analyzing the minima of $f$, 
one can focus on $p\geq 0$. The effects of the interband pairing are described by the parameter $q$. At $q=0$, TR symmetry-breaking states are only possible for $\sigma>0$.

The critical points of $f$ are found from the equations
\begin{equation}
\label{f-critical-points}
  \left\{ \begin{array}{l}
  \sigma\sin(\varphi_1+\varphi_2)-p\sin\varphi_1+q\sin(\varphi_1+\varphi_2)=0,\smallskip \\
  \sigma\sin(\varphi_1-\varphi_2)-p\sin\varphi_2-q\sin(\varphi_1+\varphi_2)=0.
  \end{array} \right.
\end{equation}
In addition to the trivial solutions $\varphi_1,\varphi_2=0$ or $\pi$, which correspond to TR invariant states, these equations can also have nontrivial solutions, in which the phases are different from $0$ and $\pi$. 
It can be shown, see Appendix \ref{app: phi1-phi2}, that all nontrivial critical points must satisfy the conditions
\begin{equation}
\label{phi1-phi2}
  \varphi_1=\varphi_2\quad\mathrm{or}\quad \varphi_1=-\varphi_2.
\end{equation}
The critical point $(\varphi_1,\varphi_2)$ is a local minimum of the free energy if the Hessian matrix 
\begin{equation}
\label{Hessian}
  \hat{\mathbb{H}}_f=\left( \begin{array}{cc}
	f_{11} & f_{12} \\
	f_{12} & f_{22}
  \end{array} \right),\quad f_{ij}=\frac{\partial^2f}{\partial\varphi_i\partial\varphi_{j}},
\end{equation}
is positive-definite, i.e., if $f_{11}>0$ and $f_{11}f_{22}-f_{12}^2>0$. The stability analysis of the critical points is done below, separately for $J>0$ and $J<0$.

\subsection{$J>0$}
\label{sec: J-positive}

\begin{figure}
\includegraphics[width=6cm]{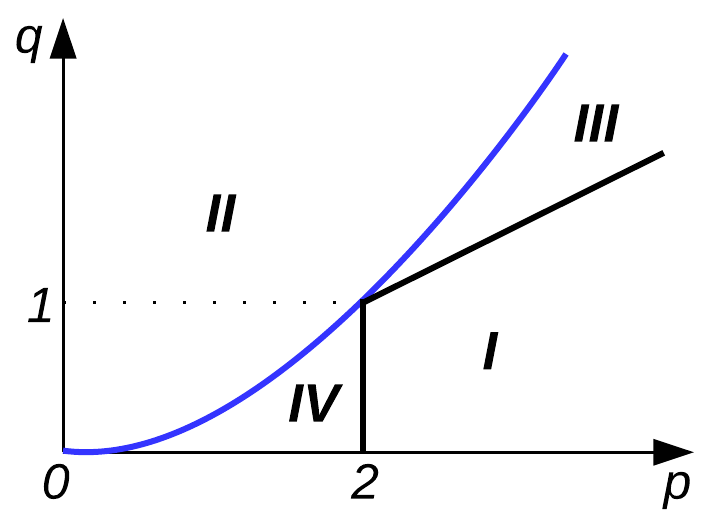}
\caption{Stable states of the free energy (\ref{f-phase-pm}) for $J>0$. The states I and II are TR invariant, while the states III and IV break TR symmetry. The black lines correspond to the continuous phase transitions between
isolated stable minima. At the blue line, the minima of Eq. (\ref{f-phase-pm}) are infinitely degenerate, as shown in Figs. \ref{fig: critical-line-1} and \ref{fig: critical-line-2}.}
\label{fig: phase-diagram-J-positive}
\end{figure}

According to the discussion above, it is sufficient to consider the following four critical points of $f$, at $p\geq 0$ and $q\geq 0$:
$$
  \left. \begin{array}{ll}
	    \mathrm{I}:\ & \varphi_1=\varphi_2=\pi,\ f_{\mathrm{I}}=1-2p+q,\\
	    \mathrm{II}:\ & \varphi_1=0,\ \varphi_2=\pi,\ f_{\mathrm{II}}=-1-q,\\
	    \mathrm{III}:\ & \varphi_1=\varphi_2,\ \cos\varphi_1=-\dfrac{p}{2q},\ f_{\mathrm{III}}=1-q-\dfrac{p^2}{2q},\\
	    \mathrm{IV}:\ & \varphi_1=-\varphi_2,\ \cos\varphi_1=-\dfrac{p}{2},\ f_{\mathrm{IV}}=-1+q-\dfrac{p^2}{2}.
  \end{array} \right.
$$
Here we also listed the energies at the critical points. The superconducting states I and II are TR invariant, whereas the states III and IV break TR symmetry. Calculating the Hessian matrix (\ref{Hessian}), 
we obtain the conditions for the critical points to be local minima of the free energy (\ref{f-phase-pm}):
\begin{equation}
\label{states-J-positive-stability}
  \left. \begin{array}{ll}
	    \mathrm{I}:\ & p>2,\ 0\leq q<\dfrac{p}{2},\\
	    \mathrm{II}:\ & q>\dfrac{p^2}{4},\\
	    \mathrm{III}:\ & p>2,\ \dfrac{p}{2}<q<\dfrac{p^2}{4},\\
	    \mathrm{IV}:\ & 0<p<2,\ 0\leq q<\dfrac{p^2}{4}
  \end{array} \right.
\end{equation}
Since the regions of local stability for different states do not overlap, there is only one stable state at each $p$ and $q$, see the phase diagram in Fig. \ref{fig: phase-diagram-J-positive}. The only exception is at $q=p^2/4$, when the 
free energy (\ref{f-phase-pm}) has a whole line of degenerate minima, instead of isolated critical points, see Appendix \ref{app: phi1-phi2}.

\begin{figure}
\includegraphics[width=6cm]{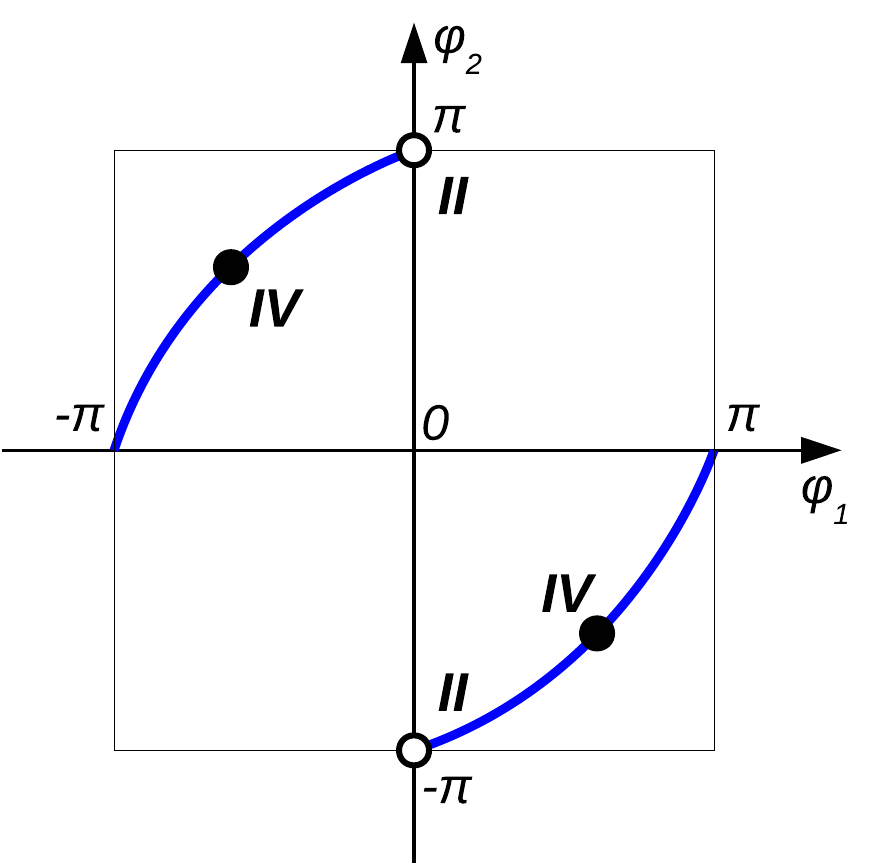}
\caption{The line of continuously degenerate minima of the free energy (\ref{f-phase-pm}) at $q=p^2/4$ and $p<2$. The empty and filled circles show the states II and IV, respectively. }
\label{fig: critical-line-1}
\end{figure}

\begin{figure}
\includegraphics[width=6cm]{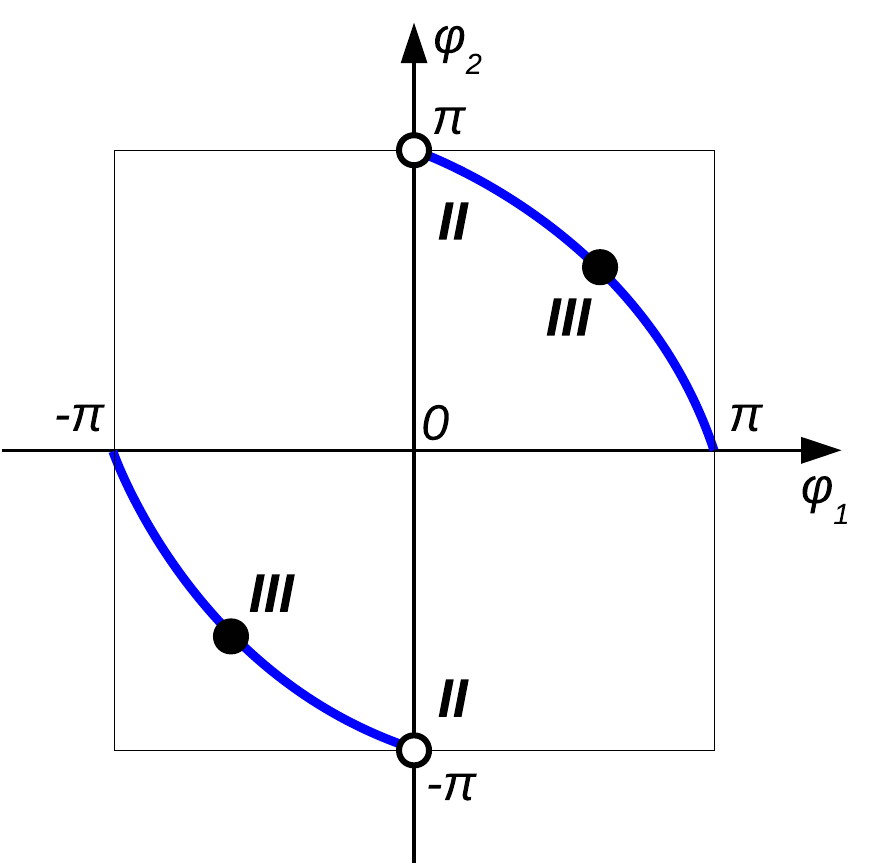}
\caption{The line of continuously degenerate minima of the free energy (\ref{f-phase-pm}) at $q=p^2/4$ and $p>2$. The empty and filled circles show the states II and III, respectively.}
\label{fig: critical-line-2}
\end{figure}

The state I is continuously transformed into the states III and IV at $q=p/2$ and $p=2$, respectively. Regarding the transition line $q=p^2/4$, the states II and IV are both located on this line at $p<2$, 
see Fig. \ref{fig: critical-line-1}, while the states II and III are both located on this line at $p>2$, see Fig. \ref{fig: critical-line-2}. Although the infinite degeneracy of the ground states at $q=p^2/4$ can possibly 
have interesting implications for the critical behaviour, we leave investigating those for a future work.

\subsection{$J<0$}
\label{sec: J-negative}

\begin{figure}
\includegraphics[width=6cm]{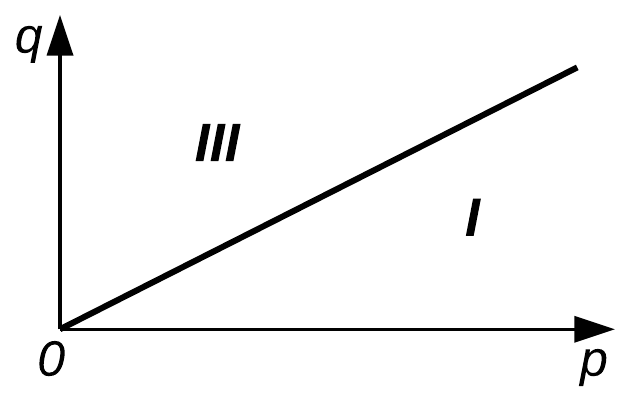}
\caption{Stable states of the free energy (\ref{f-phase-pm}) for $J<0$. The state I is TR invariant, while the state III breaks TR symmetry. }
\label{fig: phase-diagram-J-negative}
\end{figure}

It is sufficient to consider the following four critical points of $f$, at $p\geq 0$ and $q\geq 0$:
$$
  \left. \begin{array}{ll}
	    \mathrm{I}:\ & \varphi_1=\varphi_2=\pi,\ f_{\mathrm{I}}=-1-2p+q,\\
	    \mathrm{II}:\ & \varphi_1=0,\ \varphi_2=\pi,\ f_{\mathrm{II}}=1-q,\\
	    \mathrm{III}:\ & \varphi_2=\varphi_1,\ \cos\varphi_1=-\dfrac{p}{2q},\ f_{\mathrm{III}}=-1-q-\dfrac{p^2}{2q},\\
	    \mathrm{IV}:\ & \varphi_2=-\varphi_1,\ \cos\varphi_1=-\dfrac{p}{2},\ f_{\mathrm{IV}}=1+q+\dfrac{p^2}{2}.
  \end{array} \right.
$$
Calculating the Hessian matrix (\ref{Hessian}), we obtain that the states II and IV do not correspond to minima of the free energy (\ref{f-phase-pm}) at $p,q\geq 0$, and that the stability conditions for the 
other two states are given by
\begin{equation}
\label{states-J-negative-stability}
  \left. \begin{array}{ll}
	    \mathrm{I}:\ 0\leq q<\dfrac{p}{2},\\
	    \mathrm{III}:\ q>\dfrac{p}{2},
  \end{array} \right.
\end{equation}
as shown in Fig. \ref{fig: phase-diagram-J-negative}. At the transition line $q=p/2$, the TR invariant superconducting state I is continuously transformed into the TR symmetry-breaking state III.

The results for the three-band model with intraband-only pairing are reproduced if one puts $q=0$. We see that even a small nonzero $q$, corresponding to a small nonzero $\tilde J_3$ in Eq. (\ref{F-phi1-phi2}), 
produces qualitative changes in the phase diagrams, both at $J>0$ and $J<0$. In particular, if $J<0$ then the superconducting state is TR invariant at $q=0$ and all $p$, 
because of the absence of ``frustration'' in the order parameter phases. However, at any nonzero $q$, a TR symmetry-breaking state with $\varphi_1=\varphi_2\neq 0$ or $\pi$ appears in the phase diagram at sufficiently small $p$.

\section{Conclusions}
\label{sec: Conclusions}

We presented a microscopic derivation of the GL free energy in a two-band superconductor with all possible Cooper pairings between the bands. Assuming a 1D pairing channel, the order parameter has three components: the intraband ones, $\eta_1$ and $\eta_2$, which describe the pair condensates in the two bands, and also the interband one, $\tilde\eta$, which describes the pairs composed of the quasiparticles from different bands. Our expression for the GL free energy differs significantly from the previously studied three-band GL functionals with intraband-only order parameters, both in the temperature dependence of the coefficients and in the structure of the quartic terms. 

For the GL energy derivation we used the extended BCS model, in which the pairing interaction energy cutoff exceeds the band splitting, so that the Fermi surfaces in both bands are located within the same pairing shell in the momentum space. In its general form this model is characterized by six coupling constants, which is reduced to three if one neglects the interband pairing (the latter limit corresponds to the usual two-band SC model). We showed that the superconducting critical temperature $T_c$ increases in the presence of the interband pairing and is suppressed by the band splitting. Due to the large number of parameters, we had to focus on some limiting cases to achieve analytical progress. In particular, a closed-form expression for $T_c$ is available only if one assumes that the latter is the smallest energy scale in the system, much smaller than the band splitting $\Eb$.

The superconducting state that emerges immediately below $T_c$ has a real order parameter and, therefore, is TR invariant. By treating the uniform terms in the GL energy in the London approximation, we found that a variety of TR symmetry-breaking states become stable at lower temperatures. The fourth-order terms specific to the interband pairing give rise to qualitative changes in the phase diagram, compared to a three-band superconductor with intraband-only pairing.

\acknowledgments
The author is grateful to M. Sigrist and M. Holst for useful discussions.
This work was supported by a Discovery Grant 2021-03705 from the Natural Sciences and Engineering Research Council of Canada.

\appendix

\section{Symmetry of the Bloch states}
\label{app: Bloch-states}

In the presence of the electron-lattice SO coupling, the conjugate Bloch states are spinors which have both spin-up and spin-down components:
\begin{equation}
\label{Bloch-states-1-2}
  \left.\begin{array}{l}
        \langle\br|\bk,n,1\rangle=\dfrac{1}{\sqrt{\cal V}}
    \begin{pmatrix} u_{\bk,n}(\br)\\ v_{\bk,n}(\br)\end{pmatrix} e^{i\bk\br},\smallskip\\
  \langle\br|\bk,n,2\rangle=\dfrac{1}{\sqrt{\cal V}}
    \begin{pmatrix} -v^*_{\bk,n}(-\br)\\ u^*_{\bk,n}(-\br) \end{pmatrix} e^{i\bk\br},
        \end{array} \right.
\end{equation}
where ${\cal V}$ is the system volume and the Bloch factors $u_{\bk,n}(\br)$ and $v_{\bk,n}(\br)$ have the same periodicity as the crystal lattice.
The states $|\bk,n,1\rangle$ and $|\bk,n,2\rangle$ form the basis of an irreducible double-valued corepresentation (corep) of the magnetic point group of the wave vector $\bk$. 
The full symmetry group of $\bk$ is ``magnetic'', because it contains the antiunitary conjugation operation ${\cal C}$. A detailed review of magnetic groups and their coreps can be found, e.g., in 
Refs. \onlinecite{BD68} and \onlinecite{BC-book}. If the crystal point group is $\mathbb{G}$, then the magnetic group at the $\Gamma$ point is ${\cal G}=\mathbb{G}+{\cal C}\mathbb{G}$. If the $\Gamma$-point corep 
is equivalent to the spin-$1/2$ corep, then the band is called a ``pseudospin band''. In general, the Bloch states at the $\Gamma$ point do not transform as the pure spin states due to the presence of 
additional orbital factors, and we have a ``non-pseudospin'' band. 

If the $n$th band transforms at the $\Gamma$ point according to a 2D double-valued corep described by $2\times 2$ matrices $\hat{\cal D}_n(g)$, then one can construct the Bloch bases
at $\bk\neq\bm{0}$ using the following prescription:\cite{Sam19-PRB}
\begin{equation}
\label{Bloch-bases-g}
  g|\bk,n,s\rangle=\sum_{s'}|g\bk,n,s'\rangle {\cal D}_{n,s's}(g), 
\end{equation}
where $g\in\mathbb{G}$ is either a proper rotation $R$ or an improper rotation $IR$. Starting with any wave vector $\bk$ in the fundamental domain of the Brillouin zone, the expression (\ref{Bloch-bases-g}) 
defines the Bloch states at the wave vector $g\bk$. In a pseudospin band, one can put $\hat{\cal D}_n(g)=\hat D^{(1/2)}(R)$ for all $g$, where $\hat D^{(1/2)}$ is the spinor representation of rotations. 
In this case, Eq. (\ref{Bloch-bases-g}) reproduces the Ueda-Rice convention,\cite{UR85} which is widely used in theory of unconventional superconductivity. In a non-pseudospin band, the $\Gamma$-point corep 
is not equivalent to the spin-$1/2$ corep, i.e., $\hat{\cal D}_n(g)\neq\hat D^{(1/2)}(R)$ for some $g$. 

Setting $g=I$ in Eq. (\ref{Bloch-bases-g}), we have $I|\bk,n,s\rangle=p_n|-\bk,n,s\rangle$, where $p_n=\pm 1$ is the parity of the $n$th band. Since the TR operation can be written as $K={\cal C}I$, we have
\begin{equation}
\label{Bloch-functions-K}
  \left.\begin{array}{l}
   K|\bk,n,1\rangle=p_n|-\bk,n,2\rangle,\smallskip\\ 
   K|\bk,n,2\rangle=-p_n|-\bk,n,1\rangle,      
  \end{array}\right.
\end{equation}
where we used the fact that ${\cal C}^2=-1$ when acting on the spin-$1/2$ wave functions. The transformation rules (\ref{c-transform-g}) and (\ref{tilde c-c}) follow immediately from Eqs. (\ref{Bloch-bases-g}) 
and (\ref{Bloch-functions-K}).

\section{Interband $s$-wave pairing}
\label{app: s-wave-interband}

The group $\mathbf{D}_{4h}$ is generated by the rotations $C_{4z}$ and $C_{2y}$, and by the inversion $I$. The corep matrices have the form\cite{BC-book,Lax-book,Sam19-PRB} 
$$
  \left.\begin{array}{l}
        \hat{\cal D}_{\Gamma_6}(C_{4z})=\hat D^{(1/2)}(C_{4z}),\ \hat{\cal D}_{\Gamma_6}(C_{2y})=\hat D^{(1/2)}(C_{2y}),\medskip\\ 
        \hat{\cal D}_{\Gamma_7}(C_{4z})=-\hat D^{(1/2)}(C_{4z}),\ \hat{\cal D}_{\Gamma_7}(C_{2y})=\hat D^{(1/2)}(C_{2y}).
        \end{array}\right.
$$
Note that $\hat{\cal D}_{\Gamma_7}$ is not equivalent to $\hat D^{(1/2)}$, reflecting the fact that $\Gamma_7$ is a non-pseudospin corep.
In the $s$-wave pairing channel, the point-group constraint (\ref{phi-equations-g-1D}) takes the following form:
\begin{equation}
\label{phi-equations-g-s-wave}
  \hat{\cal D}_n(g)\hat\phi_{nn'}(g^{-1}\bk)\hat{\cal D}^\dagger_{n'}(g)=\hat\phi_{nn'}(\bk). 
\end{equation}
Using the expression $\hat{D}^{(1/2),\dagger}(R)\hat{\sigma}_\mu\hat{D}^{(1/2)}(R)=\sum_{\nu=1}^3R_{\mu\nu}\hat{\sigma}_\nu$, where $\hat R$ is the $3\times 3$ rotation matrix, 
we obtain that the singlet and triplet interband components satisfy the following equations:
\begin{equation}
\label{ab-equations}
  \left.\begin{array}{c}
  \tilde\alpha(\bk)=\pm\tilde\alpha(C_{4z}^{-1}\bk),\quad \tilde\alpha(\bk)=\tilde\alpha(C_{2y}^{-1}\bk),\smallskip\\
  \tilde{\bm{\beta}}(\bk)=\pm C_{4z}\tilde{\bm{\beta}}(C_{4z}^{-1}\bk),\quad \tilde{\bm{\beta}}(\bk)=C_{2y}\tilde{\bm{\beta}}(C_{2y}^{-1}\bk).
  \end{array}\right.
\end{equation}
The upper signs are realized in the $(\Gamma_6,\Gamma_6)$ or $(\Gamma_7,\Gamma_7)$ bands, while the lower signs -- in the $(\Gamma_6,\Gamma_7)$ bands.  

Some components of the interband pairing vanish identically for symmetry reasons. According to Eq. (\ref{ab-equations}), 
the $C_{2z}$ invariance constraint for $\tilde\alpha$ has the form $\tilde\alpha(\bk)=\tilde\alpha(C_{2z}^{-1}\bk)$ for all band combinations.
However, the $C_{2z}$ rotation acting on 2D wave vectors is equivalent to inversion: $\tilde\alpha(C_{2z}^{-1}\bk)=\tilde\alpha(-\bk)$. Therefore, $\tilde\alpha=0$ if the bands have opposite parity.
Similarly, we have $\tilde{\bm{\beta}}(\bk)=C_{2z}\tilde{\bm{\beta}}(C_{2z}^{-1}\bk)=C_{2z}\tilde{\bm{\beta}}(-\bk)$ for all band combinations. Therefore, $\tilde\beta_x=\tilde\beta_y=0$ if the bands have the same parity, and
$\tilde\beta_z=0$ if the bands have opposite parity.

One can easily find the lowest-order polynomial solutions of the equations (\ref{ab-equations}). For instance, for the pairs of opposite-parity bands $(\Gamma_6^\pm,\Gamma_6^\mp)$ or $(\Gamma_7^\pm,\Gamma_7^\mp)$ we have 
$\tilde\alpha=\tilde\beta_z=0$, while $\tilde{\bm{\beta}}=(\tilde\beta_x,\tilde\beta_y)$ is an odd function of $\bk$ satisfying $\tilde{\bm{\beta}}(\bk)=g\tilde{\bm{\beta}}(g^{-1}\bk)$, with $g=C_{4z}$ or $C_{2y}$. 
The simplest solution is $\tilde{\bm{\beta}}(\bk)=\bk$, which produces the gap functions (\ref{Delta12-s-wave-G6G6}). In a similar fashion, one can obtain all other expressions in Table \ref{table: phis-s-wave}.

\section{Local attractive interaction}
\label{app: V-example}

The origin of the interband pairing terms in Eq. (\ref{Hint-gen}) can be illustrated using a simple model of an attractive local interaction in a crystal without the SO coupling. In real space, we have
\begin{equation}
\label{local-model}
  \hat H_{int}=-\upsilon\int d^2\br\,\psi^\dagger_\uparrow(\br)\psi^\dagger_\downarrow(\br)\psi_\downarrow(\br)\psi_\uparrow(\br),
\end{equation}
with the coupling constant $\upsilon>0$ and the field operators given by
\begin{equation}
\label{field-operators}
  \begin{array}{l}
    \psi_\uparrow(\br)=\dfrac{1}{\sqrt{\cal V}}\sum\limits_{\bk,n}e^{i\bk\br}u_{\bk,n}(\br)c_{\bk,n1},\smallskip\\
    \psi_\downarrow(\br)=\dfrac{1}{\sqrt{\cal V}}\sum\limits_{\bk,n}e^{i\bk\br}p_nu_{\bk,n}(\br)c_{\bk,n2}.
  \end{array}
\end{equation}
Here we used the expressions (\ref{Bloch-states-1-2}) and the fact that in the absence of the SO coupling, one can put $v_{\bk,n}=0$. The lattice-periodic Bloch factors $u_{\bk,n}$ satisfy the 
symmetry relations $u^*_{\bk,n}(\br)=u_{-\bk,n}(\br)$ and $u_{\bk,n}(-\br)=p_nu_{-\bk,n}(\br)$, which, taken 
together with Eq. (\ref{tilde c-c}), make sure that $K\psi_\uparrow(\br)K^{-1}=\psi_\downarrow(\br)$ and $K\psi_\downarrow(\br)K^{-1}=-\psi_\uparrow(\br)$. 
For simplicity, we assume that the orbital wave functions at the $\Gamma$ point in both bands correspond to the identity irrep $\Gamma_1^+$ of $\mathbf{D}_{4h}$. 
This means that $p_1=p_2=1$ and, if the spin is included, then both bands correspond to the pseudospin double-valued corep $\Gamma_6^+$.

Substituting Eq. (\ref{field-operators}), neglecting the ``umklapp'' contributions, and using the identity
$$
  c^\dagger_{\bk+\bq/2,n1}\tilde c^\dagger_{\bk-\bq/2,n'1}=c^\dagger_{-\bk+\bq/2,n'2}\tilde c^\dagger_{-\bk-\bq/2,n2},
$$
the interaction Hamiltonian (\ref{local-model}) can be brought to the form (\ref{Hint-gen}) with
\begin{eqnarray}
\label{local-model-V-gen}
  && V_{s_1s_2s_3s_4}^{n_1n_2n_3n_4}(\bk,\bk';\bq)=-\frac{\upsilon}{2}\delta_{s_1s_2}\delta_{s_3s_4}\nonumber\\
  && \quad\times\left\langle u^*_{\bk+\bq/2,n_1}u_{\bk-\bq/2,n_2}u^*_{\bk'-\bq/2,n_3}u_{\bk'+\bq/2,n_4}\right\rangle_c,\qquad 
\end{eqnarray}
where $\langle(...)\rangle_c$ denotes the average over the crystal unit cell. The momentum dependence of the pairing interaction in the band representation originates from that of the Bloch factors, 
which in turn can be found using the standard $\bk\cdot\bp$ perturbation theory. The leading contributions to the pairing interaction near the $\Gamma$ point are obtained by substituting 
$u_{\bk,n}(\br)\to u_n(\br)\equiv u_{\bk=\bm{0},n}(\br)$ in Eq. (\ref{local-model-V-gen}):
\begin{equation}
\label{local-model-V-Gamma-point}
  V_{s_1s_2s_3s_4}^{n_1n_2n_3n_4}=-\frac{\upsilon}{2}\delta_{s_1s_2}\delta_{s_3s_4}\left\langle u_{n_1}u_{n_2}u_{n_3}u_{n_4}\right\rangle_c+(...).
\end{equation}
Here $u_1(\br)$ and $u_2(\br)$ are the Bloch factors, which are real and invariant under all symmetry operations from $\mathbf{D}_{4h}$, and the ellipsis stands for the momentum-dependent terms, which we neglect.

It is easy to see that the pairing interaction (\ref{local-model-V-Gamma-point}) has the factorized form (\ref{V-factorized}), with the basis functions given by
$\hat\phi_{nn'}=\hat\sigma_0$. This obviously corresponds to the $s$-wave pairing channel, with
$$
  \alpha_1(\bk)=\alpha_2(\bk)=\tilde\alpha(\bk)=1,\quad \tilde{\bm{\beta}}(\bk)=0,
$$
see Sec. \ref{sec: s-wave-D-4h}. For the coupling constants, we obtain: 
\begin{eqnarray*}
  & V_{11,11}=\upsilon\langle u_{1}^4(\br)\rangle_c,\ V_{22,22}=\upsilon\langle u_{2}^4(\br)\rangle_c,\\ 
  & V_{11,22}=\upsilon\langle u_{1}^2(\br)u_{2}^2(\br)\rangle_c,\\
  & V_{11,12}=\upsilon\langle u_{1}^3(\br)u_{2}(\br)\rangle_c,\ V_{12,22}=\upsilon\langle u_{1}(\br)u_{2}^3(\br)\rangle_c,\\ 
  & V_{12,12}=\upsilon\langle u_{1}^2(\br)u_{2}^2(\br)\rangle_c.
\end{eqnarray*}
All six independent coupling constants are nonzero and generically have the same order of magnitude. In the model (\ref{local-model}), the momenta of the band electrons are allowed to take any values in the
first Brillouin zone, so that the interaction energy cutoff is given by the bandwidth.

\section{Derivation of the GL functional}
\label{app: effective action}

We derive the free energy of the two-band superconductor with interband pairing using the effective bosonic action formalism.\cite{Popov-book} 
The starting point is the representation of the partition function in the form of a Grassmann functional integral: 
\begin{eqnarray*}
  Z &=& \mathrm{Tr}\,e^{-\beta\hat H} \\
  &=& \int{\cal D}c{\cal D}\bar c\,e^{-\int_0^\beta d\tau[\sum_{\bk,ns}\bar c_{\bk,ns}\partial_\tau c_{\bk,ns}+H_0(\tau)+H_{int}(\tau)]}, 
\end{eqnarray*}
where $\beta=1/T$. The fermionic fields $c_{\bk,ns}(\tau)$ and $\bar c_{\bk,ns}(\tau)$ are labelled by the band index $n=1,2$ and the Kramers index $s=1,2$. 
The Hamiltonian is given by Eqs. (\ref{H0-gen}) and (\ref{Hint-gen}). 

Using the factorized expression (\ref{V-factorized}) for the pairing interaction in a $d$-dimensional pairing channel, we introduce the pair fields
\begin{eqnarray*}
	& \bar B^{a}_{nn'}(\bq,\tau)=\dfrac{1}{\cal V}\sum\limits_{\bk,ss'}\phi_{nn',ss'}^{a}(\bk)\bar c_{\bk+\bq/2,ns}\tilde{\bar c}_{\bk-\bq/2,n's'},\\
	& B^{a}_{nn'}(\bq,\tau)=\dfrac{1}{\cal V}\sum\limits_{\bk,ss'}\phi_{n'n,s's}^{a,*}(\bk)\tilde c_{\bk-\bq/2,ns}c_{\bk+\bq/2,n's'},
\end{eqnarray*}
where $a=1,...,d$ and the ``time-reversed'' fermionic fields are defined in the same way as the corresponding operators, i.e., $\tilde c_{\bk,ns}(\tau)=p_n\sum_{s_1}c_{-\bk,ns_1}(\tau)(-i\hat\sigma_{y})_{s_1s}$,
see Eq. (\ref{tilde c-c}). Note that $\bar B^{a}_{nn'}=\bar B^{a}_{n'n}$ and $B^{a}_{nn'}=B^{a}_{n'n}$, according to the anticommutation condition (\ref{phi-anticommutation-constraint}). The interaction part of the action takes the following form:
$$
	S_{int}=-\frac{\cal V}{4}\int_0^\beta d\tau\sum_{\bq,a}(\bar B^a_{11},\bar B^a_{22},2\bar B^a_{12})\hat W
	\left( \begin{array}{c}
	B^a_{11} \\ B^a_{22} \\ 2B^a_{12}
	\end{array} \right),
$$
where $\hat W$ is given by Eq. (\ref{W-matrix-123}).

Next, we use the Hubbard-Stratonovich transformation to decouple the interaction part:
\begin{eqnarray*}
	&& e^{-S_{int}}\propto\int\prod_a{\cal D}^2\eta^a_{11}{\cal D}^2\eta^a_{22}{\cal D}^2\eta^a_{12}\,e^{-S_{eff,0}} \\
	&& \quad\times \exp\biggl[-\frac{{\cal V}}{2}\int_0^\beta d\tau\sum_{\bq,a}\sum_{nn'}(\eta^a_{nn'}\bar B^a_{nn'}+\eta^{a,*}_{nn'}B^a_{nn'})\biggr].
\end{eqnarray*}
Here $\eta^a_{11}(\bq,\tau)$, $\eta^a_{22}(\bq,\tau)$, and $\eta^a_{12}(\bq,\tau)=\eta^a_{21}(\bq,\tau)$ are complex bosonic fields, which can be interpreted as the fluctuating order parameter components, and 
\begin{eqnarray}
\label{S-eff-0}
	&& S_{eff,0}\nonumber\\
    && \quad={\cal V}\int_0^\beta d\tau\sum_{\bq,a}(\eta^{a,*}_{11},\eta^{a,*}_{22},\eta^{a,*}_{12})\hat W^{-1}
	\left( \begin{array}{c}
	\eta^a_{11} \\ \eta^a_{22} \\ \eta^a_{12}
	\end{array} \right).\qquad
\end{eqnarray} 
In order for the bosonic integral to be well-defined, the matrix of the coupling constants $\hat W$ has to be positive-definite. 

Introducing eight-component fermionic fields
$$
    \mathbb{C}(\bk,\tau)=(c_{\bk 11},c_{\bk 12},\tilde{\bar c}_{\bk 11},\tilde{\bar c}_{\bk 12},c_{\bk 21},c_{\bk 22},\tilde{\bar c}_{\bk 21},\tilde{\bar c}_{\bk 22})^\top,
$$
we arrive at the following expression for the partition function: 
\begin{eqnarray}
\label{Z-eta-c}
  Z &\propto& \int{\cal D}^2\bmeta\,e^{-S_{eff,0}[\bmeta^*,\bmeta]} \nonumber\\
  && \times \int{\cal D}^2c\,\exp\biggl(\frac{1}{2}\int_0^\beta d\tau\,\bar{\mathbb{C}}\hat{\cal G}^{-1}\mathbb{C}\biggr).
\end{eqnarray}
Here and below we use the shorthand notation $\bmeta$ for the set of fields $\eta^a_{11}$, $\eta^a_{22}$, and $\eta^a_{12}$. In the fermionic action, the summation over the momenta as well as over the band, Kramers, 
and Nambu indices is implied, and the Green's operator is given by
\begin{widetext}
\begin{equation}
\label{GF-1-general}
  \hat{\cal G}^{-1}_{\bk ns,\bk'n's'}=
	\left( \begin{array}{cc}
	\delta_{\bk\bk'}\biggl[-\delta_{nn'}\delta_{ss'}\dfrac{\partial}{\partial\tau}-\epsilon_{nn',ss'}(\bk)\biggr] & -\Delta_{nn',ss'}\left(\dfrac{\bk+\bk'}{2},\bk-\bk',\tau\right) \\
	-\Delta^*_{n'n,s's}\left(\dfrac{\bk+\bk'}{2},\bk'-\bk,\tau\right) & \delta_{\bk\bk'}\biggl[-\delta_{nn'}\delta_{ss'}\dfrac{\partial}{\partial\tau}+\bar\epsilon_{nn',ss'}(\bk)\biggr]
	\end{array} \right),
\end{equation}
\end{widetext}
where
\begin{eqnarray*}
  & \hat\epsilon_{nn'}(\bk)=\delta_{nn'}\xi_n(\bk)\hat\sigma_0,\\ 
  & \hat{\bar\epsilon}_{nn'}(\bk)=p_np_{n'}\hat\sigma_y\hat\epsilon^\top_{n'n}(-\bk)\hat\sigma_y=\delta_{nn'}\xi_n(\bk)\hat\sigma_0,
\end{eqnarray*}
and the dynamical gap function fields have the form
$$
	\hat\Delta_{nn'}(\bk,\bq,\tau)=\sum_{a=1}^d\eta^a_{nn'}(\bq,\tau)\hat\phi^a_{nn'}(\bk).
$$
Calculating the Grassmann integral in Eq. (\ref{Z-eta-c}), we obtain $Z\propto\int{\cal D}^2\bmeta\,e^{-S_{eff}[\bmeta^*,\bmeta]}$, where 
\begin{equation}
\label{S-eff-complete}
  S_{eff}[\bmeta^*,\bmeta]=S_{eff,0}[\bmeta^*,\bmeta]-\frac{1}{2}\mathbb{Tr}\ln\hat{\cal G}^{-1}
\end{equation}
is the effective bosonic action, with ``$\mathbb{Tr}$'' denoting the trace in the $\bk\tau$-space and the matrix trace with respect to the band, Kramers, and Nambu indices.

The order parameter components in an equilibrium superconducting state correspond to the static solutions $\bmeta(\bq)$ of the saddle-point equations $\delta S_{eff}/\delta\bmeta^*=0$. Using Eqs. (\ref{S-eff-0}) 
and (\ref{GF-1-general}), the saddle-point action has the form $S_{eff}=\beta\calF$, where  
\begin{eqnarray}
\label{F-general}
  \calF={\cal V}\sum_{\bq,a}(\eta^{a,*}_{11},\eta^{a,*}_{22},\eta^{a,*}_{12})\hat W^{-1}
	\left( \begin{array}{c}
	\eta^a_{11} \\ \eta^a_{22} \\ \eta^a_{12}
	\end{array}\right)\nonumber\\
	- \frac{1}{2}T\sum_m\Tr\ln\left(\hat{\cal G}_0^{-1}-\hat\Sigma\right)
\end{eqnarray}
is the free energy. In the second term, the summation is carried out over the fermionic Matsubara frequency $\omega_m=(2m+1)\pi T$, ``$\Tr$'' stands for the trace in the momentum, band, Kramers, and Nambu spaces,
$\hat{\cal G}_0^{-1}(\bk,\bk';\omega_m)=\delta_{\bk\bk'}\hat{\cal G}_0^{-1}(\bk,\omega_m)$, where
\begin{equation}
\label{G-0-matrix-normal}
    \hat{\cal G}_0^{-1}(\bk,\omega_m)=
    \left( \begin{array}{cc}
    \hat g_1^{-1}(\bk,\omega_m) & 0 \\
    0 & \hat g_2^{-1}(\bk,\omega_m)
    \end{array} \right) 
\end{equation}
is the inverse matrix Green's function in the normal state, with 
$$
    \hat g_n^{-1}(\bk,\omega_m)=\left( \begin{array}{cc}
         i\omega_m-\xi_n(\bk) &  0\\
         0 & i\omega_m+\xi_n(\bk)
    \end{array} \right)\otimes\hat\sigma_0,
$$
and the superconducting pairing is described by the self-energy matrix
\begin{equation}
\label{Sigma-88}
    \hat\Sigma(\bk,\bk')=\left( \begin{array}{cc}
         \hat\Sigma_{11}(\bk,\bk')  &  \hat\Sigma_{12}(\bk,\bk') \\
         \hat\Sigma_{21}(\bk,\bk')  &  \hat\Sigma_{22}(\bk,\bk')
    \end{array} \right),
\end{equation}
where
\begin{eqnarray*}
    && \hat\Sigma_{nn'}(\bk,\bk') \\
    &&   = \left( \begin{array}{cc}
         0 & \hat\Delta_{nn'}\left(\dfrac{\bk+\bk'}{2},\bk-\bk'\right) \\
         \hat\Delta_{n'n}^\dagger\left(\dfrac{\bk+\bk'}{2},\bk'-\bk\right) & 0
    \end{array} \right).
\end{eqnarray*}
The gap functions here are $2\times 2$ matrices in the Kramers space:
\begin{equation}
\label{Delta-expansion} 
  \hat\Delta_{nn'}(\bk,\bq)=\sum_{a=1}^d\eta^a_{nn'}(\bq)\hat\phi^a_{nn'}(\bk).
\end{equation}
The order parameter can be transformed into the coordinate representation as follows: $\bmeta(\br)=\sum_{\bq}\bmeta(\bq)e^{i\bq\br}$.

In a uniform superconducting state, we have $\bmeta(\bq)=\bmeta\delta_{\bq,\bm{0}}$, where $\bmeta$ is found from the self-consistency equations $\partial\calF/\partial\bmeta^*=0$. 
Taking the thermodynamic limit ${\cal V}\to\infty$, we obtain from Eq. (\ref{F-general}):
\begin{equation}
\label{gap-eq-general}
	\hat W^{-1}\left( \begin{array}{c}
	\eta^a_{11} \\ \eta^a_{22} \\ \eta^a_{12}
	\end{array}\right)=T\sum_m\int\frac{d^2\bk}{(2\pi)^2}\,
	\left( \begin{array}{c}
	\Theta^a_{11} \\ \Theta^a_{22} \\ \Theta^a_{12}
	\end{array}\right),
\end{equation}
where 
$$
  \Theta^a_{nn'}(\bk,\omega_m) = \frac{1}{2}\,\tr\biggl(\frac{\partial\hat{\cal G}^{-1}}{\partial\eta^{a,*}_{nn'}}\,\hat{\cal G}\biggr),
$$
$\hat{\cal G}^{-1}(\bk,\omega_m)=\hat{\cal G}_0^{-1}(\bk,\omega_m)-\hat\Sigma(\bk)$, and ``$\tr$'' stands for the $8\times 8$ matrix trace in the band, Kramers, and Nambu spaces.
The critical temperature $T_c$ of the phase transition into a uniform superconducting state marks the emergence of a nonzero solution of the coupled nonlinear equations (\ref{gap-eq-general}). In the case of the $s$-wave pairing, $\hat{\cal G}(\bk,\omega_m)$ can be calculated in a closed form and we arrive at the gap equations (\ref{gap-eq-s-wave-final}). 

In the general case, we retain the $\bq$-dependence of the order parameter, assume that the phase transition at temperature $T_c$ is of the second order, and expand the second term in Eq. (\ref{F-general}) in the vicinity of $T_c$ in powers of $\hat\Sigma$, i.e., in powers of the order parameter components. In this way, we obtain: $\calF=\calF_N+\calF_{GL}$, where $\calF_N$ is the normal-state free energy and $\calF_{GL}$ is the GL energy:
\begin{equation}
\label{F-GL-general}
  \calF_{GL}[\bmeta^*,\bmeta]=\calF_2^{(0)}+\calF_2^{(1)}+\calF_4+...,
\end{equation}
with $\calF_2^{(0)}$ given by the first term in Eq. (\ref{F-general}) and
\begin{eqnarray}
\label{F-GL-2-1}
  \calF_2^{(1)}=\frac{1}{2} \sum_{\bk,\bq}\sum_{n_{1,2}}
  \tr\bigl[\hat\Delta_{n_1n_2}(\bk,\bq)\hat\Delta^\dagger_{n_1n_2}(\bk,\bq)\bigr] \nonumber\\
  \times T\sum_m G_{n_1}(\bk_+,\omega_m){\bar G}_{n_2}(\bk_-,\omega_m).
\end{eqnarray}
Here ``$\tr$'' stands for the trace in the Kramers space,
\begin{equation}
\label{GFs}
  \begin{array}{c}
   G_n(\bk,\omega_m)=\dfrac{1}{i\omega_m-\xi_n(\bk)},\smallskip \\ 
  {\bar G}_n(\bk,\omega_m)=\dfrac{1}{i\omega_m+\xi_n(\bk)}=-G_n(-\bk,-\omega_m)
  \end{array}
\end{equation}
are the normal-state Green's functions, and $\bk_\pm=\bk\pm\bq/2$. The gap functions are given by Eq. (\ref{Delta-expansion}) and we obtain:
\begin{equation}
\label{F-GL-2-1-etas}
  \calF_2^{(1)}=-{\cal V}\sum_{\bq,ab}\sum_{n_{1,2}} {\cal C}^{ab}_{n_1n_2}(\bq)\eta^a_{n_1n_2}(\bq)\eta^{b,*}_{n_1n_2}(\bq),
\end{equation}
where
\begin{eqnarray}
\label{Cooperon}
   {\cal C}^{ab}_{n_1n_2}(\bq) = \frac{1}{2{\cal V}}\sum_{\bk}\tr\bigl[\hat\phi^a_{n_1n_2}(\bk)\hat\phi^{b,\dagger}_{n_1n_2}(\bk)\bigr] \nonumber\\ 
    \times T\sum_m G_{n_1}(\bk_+,\omega_m)G_{n_2}(-\bk_-,-\omega_m)\\
    ={\cal C}^{ab}_{n_1n_2}(\bm{0})+K^{ab}_{n_1n_2,ij} q_iq_j+{\cal O}(q^4)\nonumber
\end{eqnarray}
is the static pair propagator, or the Cooperon. The terms linear in $\bq$ vanish, because the basis functions have a definite parity for all band combinations, while the quadratic terms produce the GL gradient energy. 

In the fourth-order term $\calF_4$, we neglect the $\bq$-dependence of the Green's functions, as well as that of the basis functions, and obtain:
\begin{eqnarray}
\label{F-GL-4}
  \calF_4 &=& \frac{{\cal V}}{2}\sum_{\bq_i,a_i}\sum_{n_i}B^{a_1a_2a_3a_4}_{n_1n_2n_3n_4}\delta_{\bq_1+\bq_3,\bq_2+\bq_4} \nonumber\\
  && \times \eta^{a_1}_{n_1n_2}(\bq_1)\eta^{a_2,*}_{n_2n_3}(\bq_2)\eta^{a_3}_{n_3n_4}(\bq_3)\eta^{a_4,*}_{n_4n_1}(\bq_4),\qquad
\end{eqnarray}
where
\begin{widetext}
\begin{eqnarray}
  \label{B-1234}
  B^{a_1a_2a_3a_4}_{n_1n_2n_3n_4} &=& \frac{1}{2{\cal V}}\sum_{\bk}\tr\bigl[\hat\phi^{a_1}_{n_1n_2}(\bk)\hat\phi^{a_2,\dagger}_{n_3n_2}(\bk)\hat\phi^{a_3}_{n_3n_4}(\bk)\hat\phi^{a_4,\dagger}_{n_1n_4}(\bk)\bigr]
  \nonumber\\
  && \times T\sum_m G_{n_1}(\bk,\omega_m)G_{n_2}(-\bk,-\omega_m)G_{n_3}(\bk,\omega_m)G_{n_4}(-\bk,-\omega_m).
\end{eqnarray}
\end{widetext}
It is easy to check, using Eq. (\ref{K-constraint-basis-functions}), that the Cooperons satisfy ${\cal C}^{ab}_{n_1n_2}(\bq)={\cal C}^{ba}_{n_2n_1}(\bq)$ and that $B^{a_1a_2a_3a_4}_{n_1n_2n_3n_4}$ is invariant under a simultaneous cyclic permutation of the lower and upper indices. The expressions (\ref{F-GL-2-1-etas}) and (\ref{F-GL-4}) can be used to derive the GL energy for any pairing symmetry in a SC with any number of bands. In this general case, the free energy depends on $N(N+1)d/2$ order parameter components $\eta^a_{nn'}$, such that $\eta^a_{nn'}=\eta^a_{n'n}$.

\subsection{1D pairing in a two-band SC}
\label{app: GL-1D-pairing}

If the pairing corresponds to a 1D irrep of the point group, then we can drop the index $a$ and obtain from Eq. (\ref{F-GL-2-1-etas}) the following expression for the quadratic terms in the GL energy:
\begin{equation}
\label{F-GL-quadratic-1D}
  {\cal F}_2=\calF_2^{(0)}+\calF_2^{(1)}={\cal V}\sum_{\bq}\bmeta^\dagger(\bq)\hat{\cal L}(\bq)\bmeta(\bq),
\end{equation}
where
\begin{equation}
\label{L-q-matrix}
  \hat{\cal L}(\bq)=\hat W^{-1}
         -\left(\begin{array}{ccc}
          {\cal C}_{11}(\bq) & 0 & 0 \\
          0 & {\cal C}_{22}(\bq) & 0 \\
          0 & 0 & 2{\cal C}_{12}(\bq)
          \end{array}\right)
\end{equation}
and ${\cal C}_{nn'}$ are the pair propagators:   
\begin{eqnarray*}
  {\cal C}_{nn'}(\bq) &=& \frac{1}{4}\int\frac{d^2\bk}{(2\pi)^2}\,\tr\bigl[\hat\phi_{nn'}(\bk)\hat\phi^{\dagger}_{nn'}(\bk)\bigr] \\
  && \times \frac{\tanh[\xi_n(\bk_+)/2T]+\tanh[\xi_{n'}(\bk_-)/2T]}{\xi_n(\bk_+)+\xi_{n'}(\bk_-)}.
\end{eqnarray*}
The intraband basis functions are given by $\hat\phi_{nn}(\bk)=\alpha_n(\bk)\hat\sigma_0$ and the interband ones -- by Eq. (\ref{interband-phi-alpha-beta}). 

Using the band dispersions (\ref{xi-12-model}) and also Eq. (\ref{integral-cutoff}), we obtain:
$$
  {\cal C}_{nn}(\bq)=\frac{1}{4}N_F\int_{-\epsilon_c}^{\epsilon_c}\frac{d\xi}{\xi}\,\left\langle\alpha_n^2\left(\tanh\frac{\xi_+}{2T}+\tanh\frac{\xi_-}{2T}\right)\right\rangle
$$
and
\begin{eqnarray*}
  && {\cal C}_{12}(\bq)=\frac{1}{4}N_F\int_{-\epsilon_c}^{\epsilon_c}\frac{d\xi}{\xi}\\
  && \quad\times \biggl\langle g^2\left(\tanh\frac{\xi_+-\Eb/2}{2T}+\tanh\frac{\xi_-+\Eb/2}{2T}\right)\biggr\rangle,
\end{eqnarray*}
where $\xi_\pm(\bk)=\xi\pm\bmv(\bk)\bq/2$ and $\bmv=\nabla_{\bk}\xi$ is the quasiparticle velocity.
The basis functions are normalized as follows: $\langle\alpha_n^2(\bk)\rangle=\langle g^2(\bk)\rangle=1$. To extract from the above expressions the contributions that logarithmically diverge at $\epsilon_c\to\infty$, 
we subtract and add their values for $\bq=\bm{0}$ and $\Eb=0$:
$$
  {\cal C}_{nn'}(\bq)=N_F\ln\left(\frac{2e^{\mathbb{C}}\epsilon_c}{\pi T}\right)+[{\cal C}_{nn'}(\bq)-{\cal C}_{nn'}(\bm{0})],
$$
where $\mathbb{C}\simeq 0.577$ is Euler's constant. Due to the fast convergence, we extend the limits of the $\xi$-integration in the bracketed term here to infinity and use the identity 
\begin{eqnarray*}
  \frac{1}{4}\int_{-\infty}^\infty\frac{d\xi}{\xi}\,\left(\tanh\frac{\xi+\epsilon}{2T}+\tanh\frac{\xi-\epsilon}{2T}-2\tanh\frac{\xi}{2T}\right) \\
  =\Psi\left(\frac{1}{2}\right)-\re\,\Psi\left(\frac{1}{2}+i\frac{\epsilon}{2\pi T}\right),
\end{eqnarray*}
where $\Psi(z)$ is the digamma function.\cite{AS65} 

In this way, we obtain the following expressions for the intraband Cooperons expanded in powers of $\bq$:
\begin{eqnarray}
\label{C-nn-final}
  \frac{1}{N_F}{\cal C}_{nn}(\bq) &=& \ln\left(\frac{2e^{\mathbb{C}}\epsilon_c}{\pi T}\right) \nonumber\\
  && -\frac{7\zeta(3)}{16\pi^2T^2}\left\langle\alpha_n^2(\bmv\bq)^2\right\rangle + {\cal O}(\bq^4),\quad
\end{eqnarray}
where $\zeta(s)$ is the Riemann zeta function, with $\zeta(3)\simeq 1.20$. Similarly, for the interband Cooperon we have
\begin{eqnarray}
\label{C-12-final}
  && \frac{1}{N_F}{\cal C}_{12}(\bq) = \ln\left(\frac{2e^{\mathbb{C}}\epsilon_c}{\pi T}\right) \nonumber\\ 
  && \quad +\Psi\left(\frac{1}{2}\right)-\re\Psi\left(\frac{1}{2}-i\frac{\Eb}{4\pi T}\right)\nonumber\\
   && \quad +\frac{1}{32\pi^2T^2}\re\Psi''\left(\frac{1}{2}-i\frac{\Eb}{4\pi T}\right)\left\langle g^2(\bmv\bq)^2\right\rangle\qquad \nonumber\\
   && \quad + {\cal O}(\bq^4).
\end{eqnarray}
By symmetry, the only nonzero angular averages here are given by $\langle\alpha_n^2v_x^2\rangle=\langle\alpha_n^2v_y^2\rangle=w_n$ and 
$\langle g^2v_x^2\rangle=\langle g^2v_y^2\rangle=\tilde w$. Substituting the expressions (\ref{C-nn-final}) and (\ref{C-12-final}) into Eq. (\ref{L-q-matrix}) and taking the thermodynamic limit
${\cal V}\to\infty$, we finally obtain ${\cal F}_2=\int d^2\br\,F_2$, with the energy density given by Eq. (\ref{A-matrix}).

From Eq. (\ref{F-GL-4}), the ``uniform'' quartic terms have the following form in the coordinate representation: ${\cal F}_4=\int d^2\br\,F_4$, with the energy density
\begin{eqnarray}
\label{F-GL-quartic-1}
  F_4 &=& \frac{1}{2}B_{1111}|\eta_1|^4+\frac{1}{2}B_{2222}|\eta_2|^4 \nonumber\\
  && +2B_{1112}|\eta_1|^2|\tilde\eta|^2+2B_{2221}|\eta_2|^2|\tilde\eta|^2 \nonumber\\ 
  && +B_{1212}|\tilde\eta|^4+B_{1122}(\eta_1\eta_2\tilde\eta^{*,2}+\mathrm{c.c.}),\quad
\end{eqnarray}
where 
\begin{widetext}
\begin{eqnarray*}
  B_{n_1n_2n_3n_4} &=& \frac{1}{2}\int\frac{d^2\bk}{(2\pi)^2}\,\tr\bigl[\hat\phi_{n_1n_2}(\bk)\hat\phi^{\dagger}_{n_3n_2}(\bk)\hat\phi_{n_3n_4}(\bk)\hat\phi^{\dagger}_{n_1n_4}(\bk)\bigr]\\
		   && \times T\sum_m \frac{1}{i\omega_m-\xi_{n_1}(\bk)}\frac{1}{i\omega_m+\xi_{n_2}(\bk)}\frac{1}{i\omega_m-\xi_{n_3}(\bk)}\frac{1}{i\omega_m+\xi_{n_4}(\bk)}
\end{eqnarray*}
\end{widetext}
can be calculated at the critical temperature. Under the assumptions of Sec. \ref{sec: H-int-symmetry}, it is legitimate to neglect the energy dependence of the basis functions and calculate the $\xi$-integrals 
with $\epsilon_c\to\infty$ before the Matsubara sums. For example, in $B_{1212}$ we have
\begin{eqnarray*}
  && T\sum_m\int_{-\infty}^\infty d\xi \frac{1}{(i\omega_m-\xi_1)^2}\frac{1}{(i\omega_m+\xi_2)^2} \\
  && \qquad =\frac{\pi T}{2}\sum_m\frac{\mathrm{sign}\,\omega_m}{(\omega_m-i\Eb/2)^3}=\frac{7\zeta(3)}{8\pi^2T_c^2}f_1\left(\frac{\Eb}{4\pi T_c}\right),
\end{eqnarray*}
where $f_1(x)$ is given by Eq. (\ref{f_1}). In this way, we arrive at Eq. (\ref{F-GL-quartic-final}).

\section{Properties of $\hat A(T)$}
\label{app: maximum-Tc}

The superconducting instability develops at the temperature $T_c$ at which the matrix (\ref{A-matrix}) loses positive definiteness. According to Sylvester's criterion, $\hat A(T)$ is positive-definite if and only if
its principal minors $\delta_1=A_{11}$, $\delta_2=A_{11}A_{22}-A_{12}^2$, and $\delta_3=\det\hat A$ are all positive. Let us show that, if all coupling constants are nonzero, then it is $\delta_3$ that changes sign 
first as the temperature is lowered.

Suppose that at $T_c$ we have $\delta_1=0$, while $\delta_2>0$ and $\delta_3>0$, so that the phase transition occurs into the reduced state $(\eta_1,0,0)$. It is easy to see that this is not possible, because 
$\delta_2(T_c)=-A_{12}^2<0$, i.e., a contradiction. 

Suppose now that $\delta_2=0$ at $T=T_c$, while $\delta_1>0$ and $\delta_3>0$, so that the phase transition occurs into the reduced state $(\eta_1,\eta_2,0)$. From 
Schur's formula, at all $T>T_c$ we have 
\begin{eqnarray*}
  \delta_3 &=& (A_{11}A_{22}-A_{12}^2)A_{33}\\
  && -(A_{22}A_{13}^2+A_{11}A_{23}^2-2A_{12}A_{13}A_{23}).                                    
\end{eqnarray*}
At $T\to T_c+0$, the first term on the right-hand side vanishes, so that $|A_{12}|=\sqrt{A_{11}A_{22}}$, therefore
$$
  \delta_3(T_c)=-\left[\sqrt{A_{11}}A_{23}-\sign\,(A_{12})\sqrt{A_{22}}A_{13}\right]^2<0.
$$
This contradicts the assumption that $\delta_3$ is still positive at $T_c$.

\section{Order parameter near $T_c$}
\label{app: OP-near-Tc}

The temperature dependence of the order parameter components can be found by solving the nonlinear GL equations. Minimizing the GL functional, see Eqs. (\ref{F-GL-quadratic}) and (\ref{F-GL-quartic-final}), in the uniform state, we obtain: 
\begin{equation}
\label{GL-nonlinear-eqs}
    \hat A(T)\bmeta=-\bm{Q},\qquad 
    \bm{Q}=\left( \begin{array}{c}
       Q_1 \\
       Q_2 \\
       Q_3
    \end{array} \right),
\end{equation}
where $\hat A$ is given by Eq. (\ref{A-matrix}) and 
\begin{eqnarray*}
    && Q_1=(2\beta_1|\eta_1|^2+\tilde\beta_1|\tilde\eta|^2)\eta_1+\tilde\beta_4\tilde\eta^2\eta_2^*,\\
    && Q_2=(2\beta_2|\eta_2|^2+\tilde\beta_2|\tilde\eta|^2)\eta_2+\tilde\beta_4\tilde\eta^2\eta_1^*,\\
    && \tilde Q=(\tilde\beta_1|\eta_1|^2+\tilde\beta_2|\eta_2|^2+2\tilde\beta_3|\tilde\eta|^2)\tilde\eta+2\tilde\beta_4\eta_1\eta_2\tilde\eta^*.
\end{eqnarray*}
Above the second-order superconducting transition at $T_c$, Eq. (\ref{GL-nonlinear-eqs}) has only the trivial solution $\bmeta=\bm{0}$.

We denote the eigenvalues of the real symmetric matrix $\hat A$ by $\alpha_0(T)$, $\alpha_1(T)$, and $\alpha_2(T)$, with the corresponding real eigenvectors $\bm{v}_0$, $\bm{v}_1$, and $\bm{v}_2$ forming an orthonormal basis. Each of the eigenvectors has two ``intraband'' and one ``interband'' components, e.g.,
$\bm{v}_0=(v_{0,1},v_{0,2},\tilde v_0)^\top$.
To account for the matrix $\hat A$ losing positive definiteness at the phase transition, we assume that $\alpha_0$ changes sign at $T_c$:
\begin{equation}
\label{zero-mode-eigenvalue}
    \alpha_0(T)=a_0(T-T_c),\quad a_0>0,
\end{equation}
whereas $\alpha_1(T_c),\alpha_2(T_c)>0$. The eigenvector $\bm{v}_0$ is called the zero mode of $\hat A$.

We can expand the order parameter in the eigenbasis of $\hat A$ as follows:
\begin{equation}
\label{eta-expansion-v}
    \bmeta=\psi_0\bm{v}_0+\psi_1\bm{v}_1+\psi_2\bm{v}_2.
\end{equation}
To find the temperature dependence of the coefficients $\psi_0$, $\psi_1$, and $\psi_2$, we substitute the expansion (\ref{eta-expansion-v}) into the GL equations (\ref{GL-nonlinear-eqs}), use the orthonormality of the eigenvectors, and obtain:
$$
    \alpha_i\psi_i=-\bm{v}_i\bm{Q}(\psi_0,\psi_1,\psi_2),\quad i=0,1,2.
$$
The right-hand sides of these equations are homogeneous cubic polynomials of $\psi_0$, $\psi_1$, $\psi_2$, and their complex conjugates. Assuming $\psi_0$ to be real positive and explicitly separating the zero-mode contributions, we have
\begin{eqnarray}
\label{psi_012-equations}
    && \alpha_0\psi_0=-2{\cal B}_0\psi_0^3+(...),\nonumber\\
    && \alpha_1\psi_1=-2{\cal B}_1\psi_0^3+(...),\\
    && \alpha_2\psi_2=-2{\cal B}_2\psi_0^3+(...),\nonumber
\end{eqnarray}
where
\begin{eqnarray*}
    {\cal B}_0 &=& \beta_1v_{0,1}^4+\beta_2v_{0,2}^4+\tilde\beta_3\tilde v_{0}^4 \\
    && +(\tilde\beta_1v_{0,1}^2+\tilde\beta_2v_{0,2}^2+2\tilde\beta_4v_{0,1}v_{0,2})\tilde v_{0}^2.
\end{eqnarray*}
Other coefficients on the right-hand sides of the equations (\ref{psi_012-equations}) can be calculated in the similar fashion.

Introducing the notation $\tau=(T_c-T)/T_c$, we obtain from Eqs. (\ref{zero-mode-eigenvalue}) and (\ref{psi_012-equations}) that just below $T_c$, i.e., at $\tau\to 0^+$, the leading temperature dependence of the expansion coefficients in Eq. (\ref{eta-expansion-v}) is given by
\begin{equation}
\label{psi_0-tau}
    \psi_0=\sqrt{\frac{a_0T_c}{2{\cal B}_0}}\,\tau^{1/2},
\end{equation}
whereas
$$
    \psi_1=-\frac{2{\cal B}_1}{\alpha_1}\psi_0^3\propto\tau^{3/2},\quad 
    \psi_2=-\frac{2{\cal B}_2}{\alpha_2}\psi_0^3\propto\tau^{3/2}
$$
are much smaller than $\psi_0$. Therefore,
\begin{equation}
\label{eta-tau-12}
    \bmeta(T)=\sqrt{\frac{a_0(T_c-T)}{2{\cal B}_0}}\,\bm{v}_0,\quad T\to T_c-0.
\end{equation}
All three order parameter components depend on temperature in the way that is usual in the Landau theory of phase transitions, with their relative magnitudes determined by the zero mode of the matrix $\hat A$. Note that $\bmeta$ is real, which means that the TR symmetry is not broken in the superconducting state immediately below $T_c$.

\section{Proof of Eq. (\ref{phi1-phi2})}
\label{app: phi1-phi2}

Introducing the notations $s_{1,2}=\sin\varphi_{1,2}$, and $c_{1,2}=\cos\varphi_{1,2}$, the critical point equations (\ref{f-critical-points}) take the following form:
$$
  \left\{ \begin{array}{l}
  (q+\sigma)s_1c_2+(q-\sigma)s_2c_1=-ps_1,\smallskip \\
  (q-\sigma)s_1c_2+(q+\sigma)s_2c_1=-ps_2.
  \end{array} \right.
$$
These equations can be ``solved'' for $s_1c_2$ and $s_2c_1$, and we obtain: $s_1^2c_2^2-s_2^2c_1^2=\sigma(p^2/4q)(s_1^2-s_2^2)$, therefore $s_1^2=s_2^2$, unless $q=p^2/4$ at $\sigma=+1$ (recall that we assume $p,q\geq 0$). 
Thus, any isolated critical point of the function (\ref{f-phase-pm}) must satisfy the condition
\begin{equation}
\label{sf1-sf2}
  |\sin\varphi_1|=|\sin\varphi_2|,
\end{equation}
therefore $\varphi_2=\pm\varphi_1$ or $\varphi_2=\pm\varphi_1+\pi$. One can check using Eq. (\ref{f-critical-points}) that the second possibility is never realized for nontrivial critical points, 
i.e., for $\varphi_1$ and $\varphi_2$ other than $0$ or $\pi$. 

The condition (\ref{sf1-sf2}) can be violated if $\sigma=+1$ and $q=p^2/4$, in which case the minima of the free energy (\ref{f-phase-pm}) satisfy
$$
  \left\{ \begin{array}{l}
  \sin(\varphi_1+\varphi_2)=-\dfrac{2}{p}(\sin\varphi_1+\sin\varphi_2),\smallskip\\
  \sin(\varphi_1-\varphi_2)=-\dfrac{p}{2}(\sin\varphi_1-\sin\varphi_2).
  \end{array} \right.
$$
It is easy to see that these two equations are not independent (they are ``inverse'' of each other), so that their solutions correspond to whole lines, instead of isolated points, in the $(\varphi_1,\varphi_2)$ plane. 
These solutions are shown in Figs. \ref{fig: critical-line-1} and \ref{fig: critical-line-2}, for $p<2$ and $p>2$, respectively.

\end{document}